\documentclass[prb, twocolumn]{revtex4}

\begin{document}

\title{
A classification of symmetric polynomials of infinite variables\\
-- a construction of Abelian and non-Abelian quantum Hall states
}

\author{Xiao-Gang Wen}
\affiliation{
Department of Physics,
Massachusetts Institute of Technology,
Cambridge, MA 02139, USA
}

\author{Zhenghan Wang}
\affiliation{
Microsoft Station Q, CNSI Bldg. Rm 2237, University
of California, Santa Barbara, CA 93106 }

\date{{\small Jan. 15, 2008}}
\begin{abstract}
Classification of complex wave functions  of infinite variables is
an important problem since it is related to the classification of
possible quantum states of matter. In this paper, we propose a way
to classify symmetric polynomials of infinite variables using the
pattern of zeros of the polynomials.  Such a classification leads
to a construction of a class of simple non-Abelian quantum Hall
states which are closely related to parafermion conformal field
theories.
\end{abstract}

\maketitle

\setcounter{tocdepth}{3}
\tableofcontents

\section{Introduction}

\subsection{Functions of infinite variables}

One of the most important problems in condensed matter physics is
to understand how particles are organized in a ground state.
Almost all the low energy properties of the system are determined
by such an organization. Mathematically, the ground state of $N$
particles is described by a wave function -- a complex function of
$N$ variables $\Phi(\v r_1,\cdots, \v r_N)$,  where $\v r_i$ is
the coordinate of the $i^\text{th}$ particle. Thus, the problem to
understand the patterns of the many-particle organizations (or in
physical terms, to understand the quantum phases of many-particle
systems) is to classify the complex wave functions $\Phi(\v
r_1,\cdots,\v r_N)$ in the $N\to \infty$ limit.

Such a classification problem is one of most fundamental problems
in physics, since it determines the possible quantum phases of
many-particle systems.  Due to the success of Landau symmetry
breaking theory in describing phases and phase
transitions,\cite{L3726} for a long time, physicists believe that
phases of matter are classified by their symmetry properties.
Mathematically, this is equivalent to believing that the wave
functions are classified by their symmetry properties, for example
whether the wave function is invariant under translation $\Phi(\v
r_i) \to \Phi(\v r_i +\v a) $ or not.  Under such a belief, the
wave functions with the same symmetries are grouped into one class
and such a class represents a single phase of matter.  This is why
group theory becomes an important mathematical foundation in
physics.

However, after the discovery of fractional quantum Hall (FQH)
states,\cite{TSG8259,L8395} it was realized that symmetry is not
enough to classify all the possible organizations encoded in the
wave functions $\Phi(\v r_1,\cdots,\v r_N)$. This is because the
wave functions that describe different FQH states have exactly the
same symmetry. Thus the wave functions of FQH states contain new
kinds of organizations of particles that has nothing to do with
symmetry.\cite{Wrig,WNtop}  The new organizations of the particles
are called topological orders.

Intuitively, what is new in the FQH wave functions is that the
wave functions contain a long-range quantum
entanglement.\cite{KP0604,LWtopent} This is why the FQH wave
functions describe new states of matter that cannot be described
by symmetries.  The wave functions with long-range entanglements
and the corresponding topological orders not only appear in FQH
systems, they also appear in various quantum spin systems.
Understanding this new class of wave functions and the resulting
new states of matter is currently a very active research direction
in condensed matter
physics.\cite{KL8795,RK8876,AM8874,WWZcsp,RS9173,Wsrvb,MR9162,Wnab,BWkmat1,R9002,FK9169,RR9984,MLB9964,MS0181,Wqoslpub,K032,FKL0331,FNS0428,LWstrnet,K0538,R0678}

To gain a deeper and more precise understanding of topological
orders and the associated long-range entanglements, we need to
solve the related mathematical problem of classifying $\Phi(\v
r_1,\cdots,\v r_N)$ in the $N\to \infty$ limit. This is a
difficult problem which is not well studied in mathematics.  The
problem is not even well defined.  But this does not mean that the
problem is not important.  It is common not to have a well defined
problem when we wander into an unknown territory.  The first task
of knowing the unknown is usually to come up with a proper
definition of the problem.

In this paper, we will not attempt to classify generic complex
wave functions $\Phi(\v r_1,\cdots,\v r_N)$. We limit ourselves to
a simpler problem of trying to classify FQH states and their
topological orders.  (For a review on topological order in FQH
states, see \Ref{Wtoprev,Wen04}.) The corresponding mathematical
problem is to classify symmetric and anti-symmetry polynomials of
$N$ variables $\Phi(z_i,\cdots,z_N)$ in the $N\to \infty$ limit.
We will first try to come up with a physically meaningful and
mathematically rigorous definition of the problem.  Then we will
solve the problem in some simple cases. This leads to a class of
``simple'' (anti-)symmetric polynomials which corresponds to a
class of ``simple'' FQH states. The constructed FQH states include
both Abelian and non-Abelian FQH states.\cite{MR9162,Wnab}

\section{FQH states and polynomials}

\subsection{FQH wave functions}
\label{fqh}

First we would like to give a brief review on FQH theory.  A FQH state is a
quantum ground state of two dimensional electrons in a magnetic field. Such a
quantum state is described by a complex wave function
\begin{equation*}
 \Psi(x_1,y_1,x_2,y_2,\cdots, x_N,y_N)
\end{equation*}
where $(x_i,y_i)$ are the coordinates of the $i^\text{th}$ electron and $N$ the
total number of electrons.  In the strong magnetic field limit, if the filling
fraction $\nu$ of a FQH state is less than 1, then
all the electrons are in the lowest Landau level. In this case,
the ground state wave function has the following form
\begin{equation*}
\Psi = \Phi(z_1,\cdots,z_N)e^{-\frac{1}{4}\sum_{i=1}^N |z_i|^2}
\end{equation*}
where $z_i=x_i+\imth y_i$ and $ \Phi(z_1,\cdots,z_N)$ is a
holomorphic function of $z_i$ [\ie $ \Phi(z_1,\cdots,z_N)$ does not
depend on $z_i^*$]. Since $\Phi(z_1,\cdots,z_N)$ has no poles,
$\Phi(z_1,\cdots,z_N)$ is a polynomial of $z_i$'s.

Due to the Fermi statistics of the electrons,
$\Phi(z_1,\cdots,z_N)$ must be an antisymmetric polynomial. If we
assume the electrons to have a Bose statistics, then
$\Phi(z_1,\cdots,z_N)$ must be a symmetric polynomial.  So to
understand the phases of FQH systems is to classify antisymmetric
or symmetric polynomials.

It turns out that for every antisymmetric polynomials
$\Phi_\text{anti-sys}(z_1,\cdots,z_N)$, one can uniquely construct
a symmetric polynomials $ \Phi_\text{sys}(z_1,\cdots,z_N) =
\frac{\Phi_\text{anti-sys}(z_1,\cdots,z_N)}{\prod_{i<j} (z_i-z_j)}
$.  Thus classifying anti-symmetric polynomials and classifying
symmetric polynomials are almost identical problems.  So in this
paper, we will assume electrons to have a Bose statistics and
concentrate on classifying symmetric polynomials.

For a system of $N$ bosonic electrons, which symmetric polynomial will
represent the ground state of the system? It will depend on the interaction
between the electrons.  If the interaction potential between two electrons has
a $\del$-function form
\begin{equation}
\label{V01}
V_1(z_1,z_2)=\del(z_1-z_2)
\end{equation}
then the ground state is
described by the symmetric polynomial
\begin{equation*}
 \Phi_{1/2}=\prod_{i<j} (z_i-z_j)^2.
\end{equation*}
Such a state has a vanishing total potential energy $V_\text{tot}=0$, where
\begin{equation*}
V_\text{tot}\equiv
 \int \prod_id^2z_i\;
\Psi^*(\{z_i\}) \sum_{i<j} V(z_i,z_j) \Psi(\{z_i\}).
\end{equation*}
The vanishing of the total potential energy $V_\text{tot}$ requires
that the wave function $\Psi(\{z_i\}$ to be zero as $z_1\to z_2$.
Since the average energy $V_\text{tot}\geq 0$ for any wave
functions, the vanishing $V_\text{tot}$ for $\Phi_{1/2}$ indicates
that $\Phi_{1/2}$ is the ground state.

If the interaction potential between two electrons is given
by\cite{H8305}
\begin{equation}
\label{V02}
V_2(z_1,z_2)= v_0\del(z_1-z_2)+ v_2 \prt^2_{z_1^*} \del(z_1-z_2) \prt^2_{z_1}
\end{equation}
with $v_0>0$ and $v_2>0$, then the ground state will be
\begin{equation*}
 \Phi_{1/4}=\prod_{i<j} (z_i-z_j)^4.
\end{equation*}
For interaction \eq{V02}, the vanishing of the total potential
energy $V_\text{tot}$ not only requires that the wave function
$\Psi(\{z_i\}$ to be zero as $z_1\to z_2$, it also requires
$\Psi(\{z_i\}$ to vanish faster than $(z_1-z_2)^2$ as $z_1\to
z_2$. This means that the symmetric polynomial must have a fourth
order zero as $z_i\to z_j$.  One such polynomial is given by
$\Phi_{1/4}=\prod_{i<j} (z_i-z_j)^4$ which has the lowest total
power of $z_i$'s.

More complicated ground states can be obtained through more
complicated interactions. For example, consider the following
three-body interaction between electrons\cite{GWW9105,Wnabhalf}
\begin{align}
\label{Vpf}
&V_{Pf}(z_1,z_2,z_3)=
\\
&\cS[
v_0 \del(z_1-z_2) \del(z_2-z_3)
-v_1 \del(z_1-z_2) \prt_{z_3^*} \del(z_2-z_3) \prt_{z_3}
]
\nonumber
\end{align}
where $\cS$ is the total symmetrization operator between $z_1$, $z_2$, and
$z_3$. Such an interaction selects the following symmetric
polynomial\cite{MR9162}
\begin{equation*}
 \Phi_{Pf}=\cA \Big(
\frac{1}{z_1-z_2}
\frac{1}{z_3-z_4}\cdots
\frac{1}{z_{N-1}-z_N}
\Big)
\prod_{i<j} (z_i-z_j)
\end{equation*}
to describe the ground state (which has a vanishing total potential energy
$V_\text{tot}$).  Here $\cA$ is the total antisymmetrization operator between
$z_1,\cdots, z_N$.

Three symmetric polynomials $\Phi_{1/2}$, $\Phi_{1/4}$, and
$\Phi_{Pf}$ contain different topological orders and correspond to
three different phases of an $N$-electron system in the $N\to
\infty$ limit. They are the filling fraction $\nu=1/2$ Laughlin
state,\cite{L8395} the filling fraction $\nu=1/4$ Laughlin state,
and the filling fraction $\nu=1$ Pffafian state.\cite{MR9162} We
would like to find a classification of symmetric polynomials such
that the above three symmetric polynomials belong to three
different classes.

\subsection{Ideal Hamiltonian and zero-energy state}

The above three examples share some common properties.  The
Hamiltonians described by  the interaction potentials $V_1$,
$V_2$, and $V_{Pf}$ are all positive definite and contain
zero-energy eigenstates.  The zero-energy eigenstates of $V_1$,
$V_2$, and $V_{Pf}$ are known and are given by $\Phi_{1/2}$,
$\Phi_{1/4}$, and $\Phi_{Pf}$.  Thus $\Phi_{1/2}$, $\Phi_{1/4}$,
and $\Phi_{Pf}$ are exact ground states of the corresponding
Hamiltonians.  Since the interaction potentials are constructed
from $\del$-functions and their derivatives, the exact ground
states (the zero-energy states) for such type of potentials are
characterized by the pattern of zeros, \ie the orders of zeros of
the ground state wave function as we bring two or more electrons
together.

In this paper, we will concentrate on such ideal Hamiltonians and
their exact zero-energy ground states.  From this point of view,
classifying FQH states corresponds to classifying patterns of
zeros in symmetric polynomials. In other words, for each pattern
of zeros, we can define an ideal Hamiltonian such that the
symmetric polynomials with the given pattern of zeros will be the
zero energy ground states of the Hamiltonian. Such symmetric
polynomials will describe a phase of a FQH system, \emph{provided
that the Hamiltonian has a finite energy gap}.

Clearly, for the ideal Hamiltonians, apart from the zero-energy
ground states, other eigenstates of the Hamiltonian all have
non-zero and positive energies. But this does not imply the
Hamiltonian to have a finite energy gap.  Only when the minimal
energy of the excitations has a finite non-zero limit when
electron number approaches to infinity, does the Hamiltonian have
a finite energy gap.  So to classify the FQH states, we not only
need to classify the patterns of zeros and the associated ideal
Hamiltonians, we also need to judge if the constructed ideal
Hamiltonian has a finite energy gap or not.  At the moment, we do
not have a good way to make such a judgement.
So here, we will concentrate on classifying the patterns of zeros
and the associated zero energy states.

\section{Pattern of zeros}

\subsection{Derived polynomials and their $D_{ab}$ characterization}
\label{dPoly}

In order to classify translation invariant symmetric polynomials
of $N$ variables $\Phi(z_1,\cdots,z_N)$ in the $N\to \infty$
limit, we need to define the polynomials for any $N$.  The key in
our definition is to introduce ``local conditions''.  Those local
conditions apply to polynomials of any numbers of variables.

From the discussion in the Section \ref{fqh}, we see that one way
to implement the local condition is to let one variable to
approach another and to specify the power of the zero
\begin{equation*}
 \Phi(z_1,\cdots,z_N)|_{z_1\to z_2} \sim (z_1-z_2)^{D_{11}} .
\end{equation*}
We would like to stress that the above local condition is consistent
with the translation invariance of the polynomial.

However, $D_{11}$ does not contain all the information that is
needed to specify various interesting polynomials.  To implement
more general local conditions, we need to bring three or more
variables together and specify the patterns of
zeros.\cite{GWW9105,RR9984}

To describe the patterns of zeros in a systematic way, we obtain from  a
polynomial $\Phi(z_1,\cdots,z_N)$
another polynomial $P'$ by letting $z_1\to z_2$:
\begin{align*}
&\ \ \
\Phi(z_1,\cdots,z_N)|_{z_1\to z_2\equiv z^{(2)}}
\nonumber\\
&\sim (z_1-z_2)^{D_{11}} P'(z^{(2)}, z_3, \cdots,z_N)
\end{align*}
Here $\sim$ means equal up to a non-zero complex constant.  The value of
$D_{11}$ encodes a part of the local conditions.
Then we let $z_3\to z^{(2)}$ in $P'(z^{(2)}, z_3, \cdots,z_N)$:
\begin{equation*}
P'(z^{(2)}, z_3, \cdots,z_N) \sim (z_3-z^{(2)})^{D_{12}}
P''(z^{(3)}, z_4, \cdots,z_N) ,
\end{equation*}
where $z^{(3)}=z^{(2)}$. In this way, we obtain a new polynomial
$P''(z^{(3)}, z_4, \cdots,z_N)$.  In general we obtain
$P(\{z_i^{(a)}\})$ where $z^{(a)}$ is a type-$a$ variable obtained
by fusing $a$ $z_i$-variables together.  Note that $z_i^{(1)}=z_i$
is the original variable.  If we view $z_i=z_i^{(1)}$ as
coordinates of electrons, then $z_i^{(a)}$ are coordinates of
bound states of $a$ electrons. We will call such a bound state a
type-$a$ particle.

$P(\{z_i^{(a)}\})$ is a symmetric polynomial that is symmetric
between variables of the same type. It satisfies certain local
conditions and form a Hilbert space.
The polynomial $P(\{z_i^{(a)}\})$ is also called a
derived polynomial since it is obtained from $\Phi(\{z_i\})$ by fusing
variables together.

The general local conditions on $\Phi(\{z_i\})$ are specified by
pattern of zeros in its derived polynomial $P(\{z_i^{(a)}\})$:
\begin{align}
\label{fscnd}
&\ \ \ \
P(z_1^{(a)},z_1^{(b)},\cdots)\big|_{z_1^{(a)}\to z_1^{(b)}\equiv z^{(a+b)}}
\\
&\sim (z_1^{(a)}-z_1^{(b)})^{D_{ab}}
\t P(z^{(a+b)}, \cdots) + o( (z_1^{(a)}-z_1^{(b)})^{D_{ab}})
\nonumber
\end{align}
where $D_{ab}$ satisfy
\begin{equation}
\label{DelCon}
D_{ab}=D_{ba} \in Z, \ \ \ \ \
D_{aa}=\text{even},     \ \ \ \ \
D_{ab}\geq 0.
\end{equation}
$\{D_{ab}\}$ is the set of data that specifies the local condition that
$\Phi(\{z_i\})$'s must satisfy. Such a set of data is called a pattern of zeros.

We note that there are many different ways to fuse $a$
$z_i$-variables into a $z^{(a)}$ variable. The different ways of
fusion may lead to different derived polynomials which are
linearly independent. Here we will impose a unique-fusion
condition on the symmetric polynomial $\Phi(\{z_i\})$: \emph{The
derived polynomials obtained from different ways of fusions are
always linearly dependent,} \ie the derived polynomials form a
one-dimensional linear space. In this paper, we will study
symmetric polynomials $\Phi(\{z_i\})$ that satisfy this
unique-fusion condition and are characterized by the data
$D_{ab}$.

Not all possible choices of $\{D_{ab}\}$ are consistent.  Only
certain choices of $\{D_{ab}\}$ correspond to symmetric polynomials
$\Phi(\{z_i\})$. The key is to find those $\{D_{ab}\}$'s that
can be realized by some polynomials $\Phi(\{z_i\})$.

To get a feeling what a consistent set of $\{D_{ab}\}$ may look
like, let us consider the following symmetric polynomials (the
Laughlin state\cite{L8395})
\begin{equation}
\label{Laughlin}
 \Phi_{1/q} (\{z_i\})=\prod_{i<j} (z_i-z_j)^q
\end{equation}
where $q$ is an even integer.  Such a symmetric polynomial leads to the
following derived polynomial
\begin{align}
\label{Pq}
&\ \ \ \ P_{1/q}(\{z_i^{(a)}\})
\\
&=\Big\{
\prod_{a<b}
\big[
\prod_{i,j} (z^{(a)}_i-z^{(b)}_j)^{qab}
\big]\Big\}
\Big\{
\prod_{a}
\big[
\prod_{i<j} (z^{(a)}_i-z^{(a)}_j)^{qa^2}
\big]\Big\}
\nonumber
\end{align}
So the symmetric polynomial $\Phi_{1/q}$ is specified by
the pattern of zeros:
\begin{equation*}
 D_{ab}=qab, \ \ \ \ a,b \in \{1,2,3,\cdots \}
\end{equation*}
where $q$ is a positive even integer.

\subsection{$S_a$ characterization of polynomials}
\label{Sach}

There is another way to implement local conditions on a translation
invariant symmetric polynomial $\Phi(\{z_i\})$.  We introduce a
sequence of integers $S_a$, $a=0,1,2,\cdots$ and require that the
minimal total powers of $z_1,\cdots,z_a$ in $\Phi(\{z_i\})$ is given
by $S_a$.\cite{RR9984} (Here $S_0$ is defined to be 0.) Thus, in
addition to $\{D_{ab}\}$, we can also use $\{S_a\}$ to characterize
a symmetric polynomial. For a translation invariant symmetric
polynomial, $\Phi(0,z_2,\cdots,z_N)\neq 0$. Thus $S_1=0$.

The two characterizations, $\{D_{ab}\}$ and $\{S_a\}$, are closely
related. One way to see the relation is to put the symmetric
polynomial $\Phi(z_1,\cdots,z_N)$ on a sphere as discussed in
Appendix \ref{EleSph}.  Let $N_\phi$ be the maximum power of $z_1$
in $\Phi(z_1,\cdots,z_N)$. Then $\Phi(z_1,\cdots,z_N)$ can be put
on a sphere with $N_\phi$ flux quanta and each variable $z_i$
carries an angular momentum $J=N_\phi/2$.

From the discussion near the end of Appendix \ref{EleSph}, We find
that each type-$a$ particle described by $z_i^{(a)}$ in
$P(\{z_i^{(a)}\})$ carries a definite angular momentum, which is
denoted as $J_a$.  Since the lowest total power of
$z_1,\cdots,z_a$ is $S_a$, the minimal total $L^z$ quantum number
for those variables is $-aJ+S_a$. Therefore the angular momentum
of the $z_i^{(a)}$ variable is
\begin{equation}
\label{SaJa}
 J_a =  aJ-S_a .
\end{equation}
Since $z_i^{(1)}=z_i$, we find that $J_1=J$.

Again, according to the discussion near the end of Appendix 
\ref{EleSph}, if we fuse two variables $z^{(a)}$ and $z^{(b)}$
into $z^{(a+b)}$, the type-$(a+b)$ particle described by
$z^{(a+b)}$ will carry an angular momentum
\begin{equation}
\label{JJJDel}
 J_{a+b}=J_a+J_b-D_{ab} .
\end{equation}

We see that $D_{ab}$ can be expressed in terms of $S_a$:
\begin{equation}
\label{DabSa}
 D_{ab}=S_{a+b}-S_a-S_b .
\end{equation}
The conditions on $D_{ab}$, \eq{DelCon},
can be translated into the conditions on $S_a$:
\begin{equation}
\label{SaCon}
S_{2a}= \text{even},\ \ \ \ \ \
S_{a+b}\geq S_a+S_b.
\end{equation}

From the recursive relation $ J_{a+1}=J_a+J_1-D_{a,1} $, we find $
S_{a+1}=S_a+D_{a,1} $.  Using $S_1=0$, we see that $S_a$ can also
be calculated from $D_{ab}$:
\begin{equation}
\label{SaDa1}
 S_a=\sum_{b=1}^{a-1} D_{b,1} .
\end{equation}
Due to the one-to-one correspondence between $\{D_{ab}\}$ and $\{S_a\}$, we
will also call the sequence $\{S_a\}$ a pattern of zeros.

\subsection{Boson occupation characterization}

The symmetric polynomial $\Phi(z_1,\cdots,z_N)$
can be written as a sum of polynomials described by boson occupations
\begin{align*}
 \Phi(\{z_i\})=
\sum_{\{\t n_l\}}  C_{\{\t n_l\}} \Phi_{\{\t n_l\}}(\{z_i\}),
\end{align*}
where $\Phi_{\{\t n_l\}}$ is a boson occupation state with $\t
n_l$ bosons occupying the $z^l$ orbital. Mathematically,
$\Phi_{\{n_l\}}(\{z_i\})$ is given by
\begin{align}
\label{Phinl} \Phi_{\{n_l\}}(z_1,\cdots,z_N)&=\sum_P \prod_{i=1}^N
z_{P(i)}^{l_i} ,
\end{align}
where $P$ is a one-to-one mapping from $\{1,\cdots,N\} \to
\{1,\cdots,N\}$, $\sum_P$ is the sum over all those one-to-one
mapping, and $l_i$, $i=1,2,\cdots$, is a sequence of ordered
integers such that the number of $l$ valued $l_i$'s is $n_l$.

What kinds of boson occupations $\{\t n_l\}$ appear in the above
sum? Let us set $z_1=0$ in $\Phi(\{z_i\})$.  Since
$\Phi(0,z_2,\cdots,z_N)\neq 0$ due to the translation invariance,
there must be a boson occupation $\{\t n_l\}$ in the above sum
that contains one boson occupying the $z^{l=0}$ orbital.  Now let
us assume that a boson occupies $z^{l=0}$, and bring the second
particle $z_2$ to $0$, the minimal power of $z_2$ in
$\Phi(0,z_2,\cdots,z_N)$ is $D_{11}$:
\begin{equation*}
  \Phi(0,z_2,\cdots,z_N)\sim
z_2^{D_{11}} P_2(z_3,z_4,\cdots)+o(z_2^{D_{11}}).
\end{equation*}
So among those $\{\t n_l\}$ which have one boson occupying the
$z^{l=0}$ orbital, there must an $\{\t n_l\}$ that contains a
second boson occupying the $z^{l_2}$ orbital where
$l_2=D_{11}=S_2-S_1$. Next, let us assume that two bosons occupy
$z^{l=0}$ and $z^{l_2}$ orbitals and we bring the third particle
$z_3$ to $0$, the minimal power of $z_3$ is $D_{21}$:
\begin{equation*}
P_2(z_3,z_4,\cdots) \sim z_3^{D_{21}}
P_3(z_4,z_5,\cdots)+o(z_3^{D_{21}}).
\end{equation*}
Thus among those $\{\t n_l\}$ which have two bosons occupying the
$z^{l=0}$ and $z^{l_2}$ orbitals, there must be an $\{\t n_l\}$
that contains a third boson occupying the $z^{l_3}$ orbital where
$l_3=D_{21}=S_3-S_2$.  This way we can show that there must an
$\{\t n_l\}$ such that the $a^\text{th}$ boson occupies the
orbital $z^{l_a}$ with $l_a=S_a-S_{a-1}.$ Here $a=1,2,\cdots$ and
$l_1=0$. Let $n_l$ be the numbers of $l_a=S_a-S_{a-1}$ that
satisfy $l_a=l$.  We see that the boson occupation state
$\Phi_{\{n_l\}}(\{z_i\})$ happens to be the state with its
$a^\text{th}$ boson occupying the orbital $z^{l_a}$. This allows
us to show that $\Phi(z_1,\cdots,z_N)$ has a form
\begin{align}
\label{PhiPhinl}
 \Phi(\{z_i\})=\Phi_{\{n_l\}}(\{z_i\})
+\sum_{\{\t n_l\}}  C_{\{\t n_l\}} \Phi_{\{\t n_l\}}(\{z_i\}),
\end{align}
Or in other words
\begin{equation}
\label{PhiPhinlnz}
\< \Phi_{\{n_l\}} |\Phi\>\neq 0 .
\end{equation}

The two sequences, $\{S_a\}$ and $\{n_l\}$, have a one-to-one
correspondence. We will call $\{n_l\}$ the boson occupation
description of the pattern of zeros $\{S_a\}$.

The boson occupation distributions $\{\t n_l\}$ that appear in the
sum in \eq{PhiPhinl} satisfy certain conditions.  First the boson
occupation $\{\t n_l\}$ can be described by a pattern of zeros
$\{\t S_a\}$. Then the conditions on $\{\t n_l\}$ can be stated as
$\t S_a\geq S_a$.  So the minimal total power of $z_1,\cdots,z_a$
in $\Phi_{\{\t n_l\}}(\{z_i\})$ is $\t S_a$ which is equal or
bigger than $S_a$.

Haldane\cite{H0633} has conjectured that $\t n_l$'s in the
expression \eq{PhiPhinl} can be obtained from $n_l$ by one or many
squeezing operations. A squeezing operation is a two-particle
operation that moves one particle from the orbital $z^{l_1}$ to
$z^{l_1'}$ and the other from $z^{l_2}$ to $z^{l_2'}$, where $l_1
< l_1'\leq l_2'< l_2$, and $l_1+l_2=l_1'+l_2'$.  We can show that,
if $\t n_l$ is obtained from $n_l$ by squeezing operations, then
the minimal total power of $z_1,\cdots,z_a$ in $\Phi_{\{\t
n_l\}}(\{z_i\})$ is equal or bigger than $S_a$.  This is
consistent with the above discussion.

Let $P_{a,J_a}$ be a projection operator acting on the state $\Phi$
on sphere. $P_{a,J_a}$ projects into the subspace where $a$
particles in $\Phi$ have a total angular momentum equal to $J_a$ or
less. We see that for a symmetric polynomial $\Phi(z_1,\cdots,z_N)$
described by a pattern of zero $S_a$, it satisfies
\begin{equation*}
 P_{N,J_N}\cdots P_{3,J_3} P_{2,J_2} \Phi(z_1,\cdots,z_N)=\Phi(z_1,\cdots,z_N),
\end{equation*}
where $J_a=aJ-S_a$. This allows us to obtain
\begin{equation}
\label{PPhinlnz}
 P_{N,J_N}\cdots P_{3,J_3} P_{2,J_2} \Phi_{\{n_l\}}(z_1,\cdots,z_N)
\neq 0 ,
\end{equation}
where $n_l$ is the boson occupation description of $S_a$.

\section{The consistent conditions on the pattern of zeros} \label{ccDab}

For a translation invariant symmetric polynomial $\Phi(\{z_i\})$,
the corresponding pattern of zeros $\{D_{ab}\}$ and $\{S_a\}$
satisfies some special properties.  Here we would like to find
those properties as much as possible.  Those properties will be
called consistent conditions on the pattern of zeros. If we find
all the consistent conditions, a set of integers $\{D_{ab}\}$ or
$\{S_a\}$ that satisfies those consistent conditions will
correspond a translation invariant symmetric polynomial
$\Phi(\{z_i\})$. We have already find some consistent conditions
\eq{SaCon} and \eq{PPhinlnz} on $\{S_a\}$. Here we would like to
find more conditions.

\subsection{Concave condition}

If we fix all variables $z_i^{(a)}$ except $z_1^{(a)}$, then the
derived polynomial $P_\al(\{z_i^{(a)}\})$ gives us a complex
function $f(z_1^{(a)})$.  The complex function $f(z_1^{(a)})$ has
isolated zeros at $z_i^{(b)}$'s and possibly also at some other
points.

\begin{figure}[tb]
\centerline{
\includegraphics[scale=0.4]{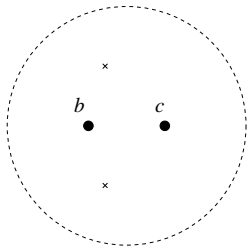}
}
\caption{
$W_{a,bc}$ obtained by moving $z_1^{(a)}$ along a large loop around
$z_1^{(b)}$ and
$z_1^{(c)}$ counts the total numbers of zeros of $f(z_1^{(a)})$ in the loop.
The crosses mark the zeros of $f(z_1^{(a)})$ not at $z_1^{(b)}$ and
$z_1^{(c)}$.
}
\label{Wabc}
\end{figure}

Let us move $z_1^{(a)}$ around two points $z_1^{(b)}$ and $z_1^{(c)}$. The
phase of the complex function $f(z_1^{(a)})$ will change by $2\pi W_{a,bc}$
where $W_{a,bc}$ is an integer (see Fig. \ref{Wabc}).  Since $f(z_1^{(a)})$
has an order $D_{ab}$ zero at $z_1^{(b)}$ and an order $D_{ac}$ zero at
$z_1^{(c)}$, the integer $W_{a,bc}$ satisfy
\begin{equation*}
 W_{a,bc} \geq D_{ab}+D_{ac} .
\end{equation*}
because $f(z_1^{(a)})$ has no poles.  Now let $z_1^{(b)}\to
z_1^{(c)}$ to fuse into $z^{(b+c)}$.  In this limit $W_{a,bc}$
becomes the order of zeros between $z_1^{(a)}$ and $z^{(b+c)}$:
$W_{a,bc}=D_{a,b+c}$. Thus we obtain the following conditions on
$D_{ab}$
\begin{equation}
\label{conCnd}
 D_{a,b+c} \geq D_{ab}+D_{ac}
\end{equation}
The concave condition \eq{conCnd} is equivalent to a condition on $S_a$:
\begin{equation}
\label{conCndW}
S_{a+b+c} +S_a+S_b+S_c \geq
S_{a+b}+S_{b+c}+S_{a+c} .
\end{equation}
We note that the Laughlin state $\Phi_{1/q} (\{z_i\})=\prod_{i<j} (z_i-z_j)^q$
saturates the above conditions: $D_{a,b+c} = D_{ab}+D_{ac}$ or $S_{a+b+c}
+S_a+S_b+S_c = S_{a+b}+S_{b+c}+S_{a+c} $.

\subsection{Symmetry condition}

\begin{figure}[tb]
\centerline{
\includegraphics[scale=0.4]{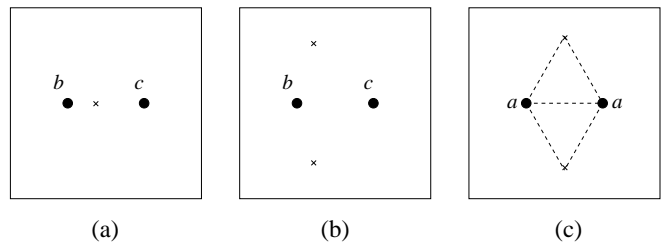}
} \caption{ Pattern of zeros of $f( z_1^{(a)}, z_2^{(b)},
z_3^{(c)})$ (when viewed as a function of $z_1^{(a)}$): (a)
$D_{a,b+c}-D_b-D_c=1$, (b) $D_{a,b+c}-D_b-D_c=2$, and (c) $a=b=c$
and $D_{a,a+a}-D_a-D_a=2$. The dash lines form two equilateral
triangles. The zeros that are not located at any variable are marked
by crosses. } \label{SWabc}
\end{figure}

If we fix all variables $z_i^{(a)}$ except $z_1^{(a)}$, $z_2^{(b)}$, and
$z_3^{(c)}$, then the derived polynomial $P_\al(\{z_i^{(a)}\})$ gives us a
complex function $f( z_1^{(a)}, z_2^{(b)}, z_3^{(c)})$.  Let us assume
$z_1^{(a)}$, $z_2^{(b)}$, and $z_3^{(c)}$ are very close to each other and far
away from all other variables.
$f(z_1^{(a)}, z_1^{(b)}, z_1^{(c)})$ has a $D_{ab}^\text{th}$
order zero as $ z_1^{(a)}\to z_2^{(b)} $ and a $D_{ac}^\text{th}$
order zero as $z_1^{(a)}\to z_3^{(c)}$.  Thus $D_{a,b+c} -
D_{ab}-D_{ac} $ is the number of zeros of $f(z_1^{(a)})$ in the
same neighborhood that are not at $z_2^{(b)}$ and $z_3^{(c)}$ (see
Fig. \ref{SWabc}).

Here we would like to assume that $\Phi(\{z_i\})$ satisfies the
unique-fusion condition.  In this case, the type-$a$ variables in
the derived polynomials have ``no shapes'' and can be treated as
points.
Since other variables are far away, the zeros of $f(z_1^{(a)},
z_2^{(b)}, z_3^{(c)})$ must satisfy certain symmetry conditions
(see Fig. \ref{SWabc}).

We see that when $a=b=c$, $f( z_1^{(a)}, z_2^{(a)}, z_3^{(a)})$ can
be zero only when $z_i^{(a)}\to z_j^{(a)}$ or when $z_1^{(a)}$,
$z_2^{(a)}$, and $z_3^{(a)}$ form an equilateral triangle. Thus the
zeros of $f( z_1^{(a)}, z_2^{(a)}, z_3^{(a)})$ (when viewed as a
function of $z_1^{(a)}$) that are marked by the crosses must appear
in pairs. We find that $D_{a,a+a}-D_a-D_a$ must be even, or
equivalently
\begin{equation}
\label{symmcond}
 S_{3a}-S_a=\text{even} .
\end{equation}

\subsection{$n$-cluster condition}
\label{secClu}

The structure of symmetric polynomials with infinite variables is
very complicated and hard to manage.  Here we would like to
introduce an $n$-cluster condition that makes a polynomial with
infinite variables behave more like a polynomial with a finite
number of variables.  A symmetric polynomial satisfies the
$n$-cluster condition if, after we fuse the variables of $\Phi(\{
z_i\})$ into $n$-variable clusters, the derived polynomial
\begin{equation}
\label{Pc}
P_c(z^{(n)}_1,\cdots,z_{N_c}^{(n)})
\sim \prod_{i<j} (z^{(n)}_i-z^{(n)}_j)^l
\end{equation}
has a simple Laughlin form where $l$ is a positive integer.

To see the structure of cluster form more clearly, let us assume
that the polynomial $\Phi(z_1,\cdots,z_N)$ describes a FQH state
with filling fraction $\nu$. This means that as a homogenous
polynomial, the total order of the $z_i$, $S_N$, in
$\Phi(z_1,\cdots,z_N)$ satisfies
\begin{equation*}
 S_N= \frac{1}{2\nu} N^2 + O(N) .
\end{equation*}
This motivates us to
write $\Phi(z_1,\cdots,z_N)$ as
\begin{align}
\label{PhiFPhi}
 \Phi(\{z_i\}) &= G(\{z_i\}) \Phi_{\nu}(\{z_i\})
\nonumber\\
\Phi_{\nu}(\{z_i\}) &= \prod_{i<j} (z_i-z_j)^{\nu^{-1}}  .
\end{align}
Here $G(\{z_i\})$ satisfies
\begin{equation*}
 G(\la z_1,\cdots,\la z_N)=\la^{s_N}G(z_1,\cdots, z_N),\ \ \ \ \ \ \ \
s_N=O(N) .
\end{equation*}
Note that $G(\{z_i\})$ is in general not a single valued function, since $
\Phi_{\nu}(\{z_i\})$ is in general not single valued.  But the product of
$G(\{z_i\})$ and $\Phi_{\nu}(\{z_i\})$ is a single-valued symmetric
polynomial.

We can fuse the variables in $G(\{z_i\})$ to obtain a derived function
$G(\{z^{(a)}_i\})$ [just as how we obtain the derived polynomial
$P(\{z_i^{(a)}\})$ from the original symmetric polynomial $\Phi(\{z_i\})$].
Similarly, we can also fuse the variables in $\Phi_\nu(\{z_i\})$ to obtain a
derived function $\Phi_\nu(\{z^{(a)}_i\})$:
\begin{equation}
\label{PhinuD}
\Phi_\nu(\{z^{(a)}_i\})=
\prod_{i,j;a<b} (z^{(a)}_i-z_j^{(b)})^{ab/\nu}
\prod_{i<j;a} (z^{(a)}_i-z_j^{(a)})^{a^2/\nu} .
\end{equation}
Thus the derived polynomial
$P(\{z_i^{(a)}\})$ can be expressed as
\begin{align}
\label{PhiFPhiD}
 P(\{z_i^{(a)}\}) &= G(\{z_i^{(a)}\})
\Phi_{\nu}(\{z_i^{(a)}\}).
\end{align}
\eq{PhiFPhiD} can be viewed as a definition of $G(\{z_i^{(a)}\})$.

Assuming $\Phi(\{z_i\})$ has an $n$-cluster form, then if we fuse
the variables of $G(\{ z_i\})$ into $n$-variable clusters, the
derived function
\begin{equation*}
 G(\{z^{(n)}_i\})=1 .
\end{equation*}
Here we will require $G(\{z^{(a)}_i\})$ to satisfy more strict
conditions
\begin{align}
\label{nClstCond}
 G(\{z^{(a)}_i\})
&= G(\{z^{(a\% n)}_i\}),
\nonumber\\
G(\cdots; z_i^{(n)};\cdots)&= G(\cdots, \cdots)
\end{align}
The second condition states that $G(\cdots,z_i^{(a)},\cdots)$ does
not depend on $z_i^{(a)}$ if $a\% n=0$. If $G$ satisfies
\eq{nClstCond}, we will say the corresponding symmetric polynomial
$\Phi(\{z_i\}) = G(\{z_i\}) \Phi_{\nu}(\{z_i\})$ to have an
$n$-cluster form.

For a symmetric polynomial $\Phi(\{z_i\})$ of an $n$-cluster form,
its pattern of zeros $D_{ab}$ can be written as
\begin{equation}
\label{Dabdab}
 D_{ab}=\nu^{-1} ab + d_{ab}
\end{equation}
where $d_{ab}$ satisfy
\begin{align}
\label{dn}
 d_{ab}&=d_{ba}
\nonumber\\
 d_{ab}&=0 \text{ if } b\% n =0
\nonumber\\
 d_{a,b+n}&=d_{ab}
\end{align}
The pattern of zeros $D_{ab}$ that satisfies the above conditions
is said to have an $n$-cluster form. Note that $\nu^{-1} ab$ in
\eq{Dabdab} describe the pattern of zeros in the derived function
$\Phi_\nu(\{z^{(a)}_i\})$ (see \eq{PhinuD}) and $d_{ab}$ describe
the pattern of zeros in the derived function $G(\{z_i^{(a)}\})$.

Setting $(a,b)=(n,n)$ and $(a,b)=(1,n)$ in \eq{Dabdab}, we find that
\begin{equation*}
 \nu^{-1} n^2 =\text{ even },\ \ \ \ \ \ \
 \nu^{-1} n =\text{ integer } .
\end{equation*}
or
\begin{equation}
\label{mneven}
 \nu^{-1} =\frac{m}{n},\ \ \ \ \ mn = \text{even} .
\end{equation}
We also find that
\begin{equation}
\label{DelModn}
 D_{a,b+n}=D_{a,b}+am
\end{equation}

Let
\begin{equation}
\label{saSa}
 s_a=S_a-\frac{1}{2\nu} a(a-1)
\end{equation}
We find that (see \eq{DabSa})
\begin{equation}
\label{dabwawb}
 d_{ab}=s_{a+b}-s_a-s_b
\end{equation}
The cluster conditions \eq{dn} become
\begin{align}
& s_{a+n}-s_a-s_n=0 ,
\nonumber\\
& s_{a+n}-s_{a}=s_{b+n}-s_{b} .
\end{align}
Since $S_1=s_1=0$ (see \eq{SaJa}), we find that $s_{n+1}=s_n$ and
$s_{a+n}-s_a=s_n$.
Thus
\begin{equation}
\label{anysa}
 s_{a+kn}=ks_n+s_a,\ \ \ \ \ a=1,2,\cdots,\infty .
\end{equation}
This allows us to obtain
$s_a$ for any $a>0$ from $s_1,s_2,\cdots,s_n$.
Similarly, all the $S_a$'s can be determined from
$S_1$, $S_2$, $\cdots$, $S_n$:
\begin{align}
\label{Sbany}
S_{a+kn}&= s_{a+kn}+\frac{m}{2n} (a+kn)(a+kn-1)
\nonumber\\
&=ks_n+s_a +\frac{m}{2n} (a+kn)(a+kn-1)
\nonumber\\
&=S_a-\frac{m}{2n}a(a-1)+ k [S_n -\frac{m}{2} (n-1)]
\nonumber\\
&\ \ \ \ \ \ \ \ \ \ \ \ \
+\frac{m}{2n} (a+kn)(a+kn-1 )
\nonumber\\
&=
S_a+kS_n +\frac{k(k-1)nm}{2} +kma
\end{align}
The above result is actually valid for
any positive integer $a$.

It is convenient to introduce
\begin{align}
\label{hscaSa}
 h^\text{sc}_a &\equiv s_a-\frac{a}{n}s_n
= S_a- \frac{aS_n}{n}+\frac{am}{2}-\frac{a^2m}{2n}.
\end{align}
From \eq{anysa}, we can show that
$h^\text{sc}_a$ is periodic:
\begin{equation*}
h^\text{sc}_{a}=h^\text{sc}_{a+n}.
\end{equation*}
Since $s_1=0$, we see that $h^\text{sc}_1=-s_n/n$ and
$ s_a=h^\text{sc}_a-ah^\text{sc}_1$.
From \eq{saSa}, we see
that $S_a$ can be calculated from $h^\text{sc}_a$:
\begin{equation}
\label{Sahsca}
 S_a=h^\text{sc}_a-ah^\text{sc}_1+\frac{a(a-1)m}{2n} .
\end{equation}
\eq{hscaSa} and \eq{Sahsca} imply that the two sequences of numbers,
 $\{S_a\}$ and
$\{h^\text{sc}_a\}$, have a one-to-one correspondence and can
faithfully represent each other.  In this paper, we will use both
sequences to characterize the symmetric polynomials.  The
$h^\text{sc}_a$ characterization turns out to have a close
relation to the CFT description of the FQH states (see Section
\ref{cft}).\cite{MR9162,BWnab,WWopa,WWHopa}

If $S_a$ has the $n$-cluster form \eq{Sbany}, then the corresponding boson
occupation numbers $n_l$ have some nice properties.  From \eq{Sbany}, we see
that $ S_{a+n}-S_{a-1+n}= S_{a}-S_{a-1}+m$. Thus $l_i=S_i-S_{i-1}$ satisfy
$l_{i+n}=l_i+m$.  This means that the boson occupation in the orbitals $z^l$
has a periodic structure: every time we skip $n$ boson, we skip $m$ orbitals.
Or in other words, if we know the occupation distribution of the first $n$
bosons, the occupation distribution of second $n$ bosons can be obtained from
that of first $n$ bosons by shifting the orbital index $l$ by $m$. Thus the
occupation numbers $n_l$ satisfy $n_l=n_{l\% m}$ (see \eq{nlm}).  Also, each
$m$ orbitals contain $n$ bosons.  Due to the one-to-one correspondence between
$S_a$ and $n_l$, we can also use $n_0,\cdots,n_{m-1}$ to describe the pattern
of zeros in $\Phi(\{z_i\})$.

\subsection{Translation invariance}

To study the translation invariance of the symmetric polynomial
$\Phi(z_1,\cdots,z_N)$, let us put the polynomial on a sphere (see
Section \ref{EleSph}) and study its rotation invariance.  In fact,
in this paper, when we mention translation invariance we actually
mean rotation invariance on sphere.

Let $N_\phi$ be the number of flux quanta going through the
sphere.  Then, each variable $z_i$ in $\Phi(z_1,\cdots,z_N)$
carries an angular momentum $J=N_\phi/2$.  What is the total
angular momentum of $\Phi(z_1,\cdots,z_N)$? In general,
$\Phi(z_1,\cdots,z_N)$ does not carry a definite angular momentum.
So here we will calculate the maximum angular momentum of
$\Phi(z_1,\cdots,z_N)$ from the pattern of zeros $D_{ab}$.

The maximum angular momentum is nothing but the angular momentum of
$z^{(N)}$ -- the particle obtained by fusing all the $N$ electrons together.
The angular momentum of $z^{(N)}$
is given by (see \eq{SaJa} and \eq{SaDa1})
\begin{equation}
\label{Jtot}
 J_{N}=J_\text{tot}= N J-\sum_{a=1}^{N-1} D_{a,1}=NJ-S_N
\end{equation}
If $J_N=0$ for a symmetric polynomial $\Phi(\{z_i\})$,
then $\Phi(\{z_i\})$ is invariant under the $O(3)$ rotation of the sphere.
In other words, $\Phi(\{z_i\})$ is translation invariant.

However, for an arbitrary choice of $N$ and $J$, $J_N$ is not zero
in general. $J_N$ can be zero only for certain combinations of
$(N,J)$.  For the filling fraction $\nu=1/q$ Laughlin state,
$J_{N}=NJ-q\frac{N(N-1)}{2}$. We find $J_{N}=0$ if
\begin{equation}
\label{NphiNq}
 2J=N_\phi=qN - q  .
\end{equation}
This is the relation between the number of magnetic flux quanta, $N_\phi$, and
the number of electrons, $N$, of the $\nu=1/q$ Laughlin state if the Laughlin
state is to fill the sphere completely (which gives rise to a rotation invariant
state).

Assume that the symmetric polynomial $\Phi(z_1,\cdots,z_N)$ has an
$n$-cluster form described by the data $(m;S_a|_{a=1,\cdots,n})$.
If we put the polynomial on a sphere, the maximum total angular
momentum of $\Phi(z_1,\cdots,z_N)$ is given by \eq{Jtot}.
If $N=n N_c$, we find
from \eq{Sbany} that
\begin{align}
\label{SnNc}
S_{nN_c}&=S_{n+ (N_c-1) n}=
N_cS_n+\frac{mnN_c(N_c-1)}{2}
\nonumber\\
J_\text{tot} &=J n N_c-N_cS_n-\frac{mnN_c(N_c-1)}{2}
\end{align}

When  $N=n N_c$, $\Phi(z_1,\cdots,z_N)$ can give rise to a Laughlin wave
function \eq{Pc} after fusing $z_i$'s into $N_c$ $z_i^{(n)}$'s (see \eq{Pc}).
Since it is always possible to fill the sphere with the Laughlin state, this
implies that there exist an integer $2J$ to make $J_\text{tot}=0$.  Such an
integer is given by
\begin{align}
\label{NphiN}
 2J=N_\phi &= \frac{2S_{nN_c}}{nN_c}=
\frac{2S_n}{n}+m(N_c-1)
\end{align}
This requires that
\begin{equation}
\label{2Snn}
2S_n=0 \text{ mod } n  .
\end{equation}

To summarize, the $n$-cluster condition requires that if $N\% n=0$
and $(N_\phi, N)$ satisfies \eq{NphiN}, then the symmetric
polynomial $\Phi(z_1,\cdots,z_N)$ must represent a rotation
invariant state on sphere.  The existence of such rotation
invariant state requires $S_n$ to satisfy \eq{2Snn}.

\section{Construction of ideal Hamiltonians}

We have seen that the pattern of zeros in an electron wave
function $\Phi(\{z_i\})$ can be described by a set of integers
$S_2,S_3,\cdots$.  In this section, we are going to construct an
ideal Hamiltonian on sphere to realize such a kind of electron
wave function as a ground state of the Hamiltonian.

On sphere, the set of integers $S_a$ also has a very physical
meaning.  For an electron system on a sphere with $N_\phi$ flux
quanta, each electron carries an orbital angular momentum
$J=N_\phi/2$ if the electrons are in the first Landau
level.\cite{H8305} For a cluster of $a$ electrons, the maximum
allowed angular momentum is $aJ$. However, for the wave function
$\Phi(\{z_i\})$ described by $S_a$, the maximum allowed angular
momentum is only $J_a=aJ-S_a$.  The pattern of zeros forbids the
appearance of angular momenta $aJ-S_a+1, aJ-S_a+2,\cdots, aJ$ for
any $a$-electron clusters in $\Phi(\{z_i\})$.

Such a condition can be easily enforced by a Hamiltonian.  Let $P^{(a)}_S$ be
a projection operator that acts on $a$-electron Hilbert space. $P^{(a)}_S$
projects onto the subspace of $a$ electrons with total angular momenta
$aJ-S+1,\cdots, aJ$.  Now consider the Hamiltonian\cite{RR9984,R0634,SRC0760}
\begin{equation}
\label{Hideal1}
 H_{\{S_a\}}=\sum_a \sum_{a\text{-electron clusters}} P^{(a)}_{S_a}
\end{equation}
where $\sum_{a\text{-electron clusters}}$ sum over all $a$-electron
clusters. The wave function $\Phi(\{z_i\})$ with a pattern of zeros
described by $S_a$, if exist, will be the zero-energy ground state
of the above Hamiltonian.

We note that the Hamiltonian $ H_{\{S_a\}}$ is well defined for
any choice of $S_a$.  However, for a generic choice of $S_a$, the
zero-energy ground state of $H_{\{S_a\}}$ may not be the one with
a pattern of zeros described by $S_a$.  This is because when we
say the wave function $\Phi(\{z_i\})$ has a
pattern of zeros described by $S_a$, we mean two things:\\
(a) The angular momenta $aJ-S_a+1,\cdots aJ$ do not appear
for any $a$-electron clusters in $\Phi(\{z_i\})$;\\
(b) The angular momenta $aJ-S_a$ does appear
for $a$-electron clusters in $\Phi(\{z_i\})$.\\
The zero-energy ground state of $H_{\{S_a\}}$ satisfies the
condition (a).  But sometimes, we may find that the condition (a)
also implies that $aJ-S_a$ does not appear for $a$-electron
clusters in $\Phi(\{z_i\})$ for certain values of $a$.  This means
that the zero-energy ground state of $H_{\{S_a\}}$ is actually
described by a pattern of zeros $\t S_a$ which satisfy $\t S_a
\geq S_a$.  However, for a certain special set of $\{S_a\}$ that
describe the pattern of zeros of an existing symmetric polynomial,
we have $\t S_a=S_a$.
For those $S_a$, the zero-energy ground state of $H_{\{S_a\}}$
is described by the pattern of zeros of $\{S_a\}$ itself.

We have seen that, for a FQH state described by a pattern of zeros
$\{S_a\}$, a state of $a$-electron clusters has a non-zero
projection into the space $\cH_{a,S_a}$, where $\cH_{a,S_a}$ is a
space with a total angular mentum $aJ-S_a$.  However, different
positions of other electrons may lead to different images in the
space $\cH_{a,S_a}$.  Let $\cH_a$ be the subspace of $\cH_{a,S_a}$
that is spanned by those images.  In general $\cH_{a,S_a}\neq
\cH_a$.  So, in general, the zero-energy ground state of the ideal
Hamiltonian $H_{\{S_a\}}$ may not be unique.  In an attempt to
construct an ideal Hamiltonian for which the FQH state $\Phi$ is
the unique ground state, we can add additional projection
operators and introduce a new ideal Hamiltonian
\begin{equation}
\label{Hideal}
 H_{\{S_a\}}=\sum_a \sum_{a\text{-electron clusters}}
(P^{(a)}_{S_a} + P_{\bar\cH_a}) ,
\end{equation}
where $P_{\bar\cH_a}$ is a projection operator
into the space $\bar\cH_a$ and $\bar\cH_a$ is a subspace of
$\cH_{a,S_a}$ formed by vectors that is perpendicular to
$\cH_a$.

\section{Summary of general results}

In last a few sections,
we have considered a subclass of symmetric polynomials
$\Phi(\{z_i\})$ (of infinity variables) that satisfy (a) a
unique-fusion condition (see discussion in Section \ref{dPoly}),
(b) an $n$-cluster condition (see discussion in Section
\ref{secClu}), and (c) translation invariance
\begin{equation*}
\Phi(\{z_i\})=\Phi(\{z_i-z\}).
\end{equation*}
The unique-fusion condition requires that when we fuse the
variables $z_i$ together to obtain new polynomials, we will always
get the same polynomial no matter how we fuse the variables
together.  The $n$-cluster condition requires that if we fuse all
the variables $z_i$ into clusters of $n$ variables each, the
resulting polynomial of the clusters has the Jastrow form
$\prod_{i<j}(z^{(n)}_i-z^{(n)}_j)^q$.

We find that each translation invariant symmetric polynomial
$\Phi(\{z_i\})$ of the $n$-cluster form and satisfying the
unique-fusion condition is characterized by a set of non negative
integers $(m;S_2,\cdots,S_n)$. However, not all sets of non
negative integers $(m;S_2,\cdots,S_n)$ can be realized by such
symmetric polynomials. The $(m;S_2,\cdots,S_n)$ that correspond to
existing translation invariant symmetric polynomials (that satisfy
the $n$-cluster and the unique-fusion conditions) must satisfy
certain conditions.

First $m$ and $S_n$ must satisfy (see \eq{mneven} and \eq{2Snn})
\begin{align}
\label{mSn}
 & m>0, \ \ \ \ \ mn=\text{even}
\nonumber\\
& 2S_n=0 \text{ mod } n
\end{align}
From $m,S_2,\cdots,S_n$ and $S_1=0$,
we can determine $S_a$ for any $a>1$:
(see \eq{Sbany})
\begin{align}
\label{Sany}
S_{a+kn} &= S_a+kS_n +\frac{k(k-1)nm}{2} +kma
\end{align}
Those $S_a$ must satisfy (see \eq{SaCon} and \eq{conCndW})
\begin{align}
\label{SanyCnd}
\Del_2(a,a) &=\text{even},
\nonumber\\
 \Del_2(a,b)&\geq 0,
&  \Del_3(a,b,c) &\geq 0  .
\end{align}
where
\begin{align}
 \label{D2D3}
\Del_2(a,b)&\equiv  S_{a+b}-S_a-S_b,
\\
\Del_3(a,b,c) &\equiv S_{a+b+c} -S_{a+b}-S_{b+c}-S_{a+c} +S_a+S_b+S_c  .
\nonumber
\end{align}

The $S_a$'s also satisfy another condition which is harder to
describe. To describe the new condition, we first note that the
sequence $\{S_a\}$ can be encoded by another sequence of
non-negative integers $n_l$, $l=0,1,\cdots$.  To obtain $n_l$ from
$S_a$, we introduce $l_a\equiv S_a-S_{a-1}$ for $a=1,2,\cdots$.
Then $n_l$ is the number of $l_a$'s that satisfy $l_a=l$. The two
sequences, $\{S_a\}$ and $\{n_l\}$, have a one-to-one
correspondence and can faithfully represent each other.  The
number $n_l$ can be regarded as the boson occupation number that
was used to characterize FQH states in the thin cylinder
limit.\cite{SL0604,BKW0608,SY0802} $n_l$ is also used to label Jack
polynomials that describe FQH states.\cite{BH0737,BH0762}

Now let us introduce $2J+1$ orbitals $|m_z\>$, $m_z=-J,-J+1,\cdots,J-1,J$, which
form a representation of $SU(2)$ with an angular momentum $J$. (Here $2J$ is
an integer.) We can create a many-boson state $|\{n_l\}\>$ by putting $n_l$
bosons into the $m_z=l-J$ orbitals.  Then $S_a$ must be such that (see
\eq{PPhinlnz})
\begin{equation}
\label{PJtot}
 P_{N,NJ-S_N}
\cdots
 P_{3,3J-S_3}
 P_{2,2J-S_2}
 |\{n_l\}\> \neq 0 ,
\end{equation}
where $P_{a,J_a}$ is a projection operator that projects into the subspace
where any $a$ particles have a total angular momentum equal to $J_a$ or less
and $N$ is the number of particles in $|\{n_l\}\>$.

We will call $(m;S_2,\cdots,S_n)$ an $\v S$-vector and denote it
as
\begin{equation*}
 \v S=(m;S_2,\cdots,S_n)
\end{equation*}
We find that translation invariant symmetric polynomials (that
satisfy the $n$-cluster and the unique-fusion conditions) are
labeled by the $\v S$-vectors that satisfy \eq{mSn}, \eq{SanyCnd},
and \eq{PJtot}.

We would like to stress that \eq{mSn}, \eq{SanyCnd}, and
\eq{PJtot} are only necessary conditions for $(m;S_2,\cdots,S_n)$
to describe a translation invariant symmetric polynomial of the
$n$-cluster form and satisfying the unique fusion condition.  We
do not know if those conditions are sufficient or not.  Some $\v
S$-vectors that satisfy \eq{mSn}, \eq{SanyCnd}, and \eq{PJtot} may
not correspond to an existing symmetric polynomial. Also, there
may be more than one symmetric polynomials that are described by
the same $\v S$-vector that satisfy \eq{mSn}, \eq{SanyCnd}, and
\eq{PJtot}. Each such symmetric polynomial corresponds to a FQH
state. By solving  \eq{mSn}, \eq{SanyCnd}, and \eq{PJtot}, we can
obtain $(m;S_2,\cdots,S_n)$'s that correspond to the Laughlin
states, the Pffafian state,\cite{MR9162} the parafermion
states,\cite{RR9984} and many new non-Abelian states.

We also obtained some additional results.
Our numerical studies of the equations \eq{mSn} and \eq{SanyCnd}
suggest that all solutions of the equations satisfy
\begin{equation}
\label{hrefl}
 h^\text{sc}_a=h^\text{sc}_{n-a} ,
\end{equation}
where
\begin{align}
\label{hscaSa1}
 h^\text{sc}_a = S_a- \frac{aS_n}{n}+\frac{am}{2}-\frac{a^2m}{2n},
\end{align}
although we cannot derive the equation \eq{hrefl} analytically.  Such a
relation implies that
\begin{equation*}
 S_{n-a}=S_a+\frac{n-2a}{n} S_n .
\end{equation*}
$h^\text{sc}_a$'s also satisfy
\begin{equation*}
 h^\text{sc}_a=h^\text{sc}_{a\% n},\ \ \ \ \ \ \ \
 h^\text{sc}_a\geq 0 .
\end{equation*}
where $a\% n\equiv a \text{ mod } n$.

From \eq{hscaSa1}, we find  that
$(h^\text{sc}_1,\cdots, h^\text{sc}_n)$
and $(S_2,\cdots,S_n)$ have a one-to-one correspondence. They can
faithfully represent each other.
Due to the one-to-one relation between $S_a$ and $h^\text{sc}_a$, we can also
use $n$, $m$, and $h^\text{sc}_1,\cdots,h^\text{sc}_n$ to characterize the
pattern of zero in the symmetric polynomial $\Phi(\{ z_i\})$.
We will package the data in the form
\begin{equation*}
\v h=(\frac{m}{n}; h^\text{sc}_1,\cdots,h^\text{sc}_n),
\end{equation*}
and call $\v h$ an $\v h$-vector.  We see that patterns of zeros
in a symmetric polynomial can also be described by the $\v
h$-vectors.

Each symmetric polynomial described by the pattern of zeros
$\{S_a\}$ is related to a conformal field theory (CFT) generated by
simple-current operators which have an Abelian fusion rule (see
section \ref{cft}). $(h^\text{sc}_1,\cdots, h^\text{sc}_n)$ turn out
to be the scaling dimensions of those simple-current operators.
Since $\Del_3(a,b,c)$ only depends on $h^\text{sc}_a$
\begin{equation*}
\Del_3(a,b,c) = h^\text{sc}_{a+b+c}
-h^\text{sc}_{a+b}-h^\text{sc}_{b+c}-h^\text{sc}_{a+c}
+h^\text{sc}_a+h^\text{sc}_b+h^\text{sc}_c  ,
\end{equation*}
$\Del_3(a,b,c)\geq 0$ is a property of the simple-current CFT.

The condition \eq{PJtot} is hard to check.  So let us just consider \eq{mSn}
and \eq{SanyCnd}.  One class of solutions of \eq{mSn} and \eq{SanyCnd} is
given by
\begin{equation*}
h^\text{sc}_a= h^{Z_n}_a \equiv \frac{ (a\%n) [n-(a\%n)] }{n} .
\end{equation*}
This class of solutions corresponds to the $Z_n$ parafermion CFT which is
generated by simple-current operators $\psi_a$ that satisfy an Abelian fusion
rule $ \psi_a \psi_b = \psi_{a+b} $ and $\psi_0=\psi_n=1$. The scaling
dimensions of $\psi_a$ is given by the above $h^{Z_n}_a$.  The parafermion
states introduced in \Ref{RR9984} are related to such a class of solutions.  

A more general class of solutions of \eq{mSn} and \eq{SanyCnd} corresponds to
generalized parafermion CFTs.  A generalized parafermion CFT is generated
simple-current operators that have the following dimensions
\begin{equation*}
h^\text{sc}_a= h^{Z_n^{k}}_a\equiv
\frac{ (k a\%n) [n-(k a\%n)] }{n} .
\end{equation*}
Those solutions represent a new class of non-Abelian FQH states, which
will be called generalized parafermion states.

It turns out that all the solutions of \eq{mSn} and \eq{SanyCnd} are closely
related to parafermion CFTs, \ie a solution $h^\text{sc}_a$ satisfies 
\begin{equation}
\label{hschscPF}
 h^\text{sc}_a= \sum_i \frac{k_i}{2} h^{\text{PF}_i}_a \text{ mod } 1
\end{equation}
where $k_i$'s are positive or negative integers and $h^{\text{PF}_i}_a$'s are
the scaling dimensions of the parafermion operators in some parafermion CFT's
labeled by $i$.  They are given by
\begin{equation*}
 h^\text{PF}_a=h^{Z_{n'}^{k}}_a
\end{equation*}
for certain integers $k$ and $n'$, where $n'$ is a factor of $n$.

If $h^\text{sc}_a=h^{\text{PF}}_a$, then the solution corresponds to an
existing symmetric polynomial generated by a (generalized) parafermion CFT. If
$h^\text{sc}_a=\frac12 h^{\text{PF}}_a$, then the solution corresponds to the
square root of a symmetric polynomial generated by a (generalized) parafermion
CFT. Therefore, the later solution does not correspond to any existing
symmetric polynomials.  Numerical experiments suggest that the later cases
always have $\Del_3(a,b,c)=$ odd for some $a$, $b$, and $c$.  This motivates
us to introduce the following new condition
\begin{equation}
\label{D3even}
 \Del_3(a,b,c)=\text{even}
\end{equation}
to exclude those illegal cases.  The conditions \eq{mSn},
\eq{SanyCnd}, and \eq{D3even} provide an easy way to obtain $S_a$'s
that may correspond to existing symmetric polynomials.  The new
condition \eq{D3even} is a generalization of necessary conditions
$\Del_3(a,a,a)=$ even (see \eq{symmcond}), \eq{Snmeven}, and $S_a \neq 1$.

Our numerical studies suggest that the solutions of the equations
\eq{mSn}, \eq{SanyCnd} and \eq{D3even} give rise to $h^\text{sc}_a$ that
satisfy
\begin{equation*}
 h^\text{sc}_a= \sum_i k_i h^{\text{PF}_i}_a \text{ mod } 2
\end{equation*}
Those solutions also have the properties that $m=$ even and $S_a=$ even.

We also find that, for $S_a$ satisfying \eq{mSn} and \eq{SanyCnd}, the
corresponding $n_l$ is a periodic function of $l$ (for $l\geq 0$) with a
period $m$:
\begin{equation}
\label{nlm}
n_l=  n_{l\% m} .
\end{equation}
The $n_l$'s satisfy
\begin{equation}
\label{Nnl}
 \sum_{l=0}^{2J} n_l = n N_c,\ \ \ \ \ \
 \sum_{l=0}^{2J} (l-J) n_l = 0,
\end{equation}
for any $(J,N_c)$ satisfying
\begin{equation}
\label{JNcSn}
2J= \frac{2S_n}{n}+m(N_c-1)
\end{equation}

\section{General structure of the solutions}

The $\v S$-vectors that satisfy \eq{mSn} and
\eq{SanyCnd} has some general properties.
In this section we will discuss those properties.

\subsection{$n$-cluster polynomial as
$\ka n$-cluster polynomial}

Let $P(\{ z_i^{(a)} \})$ be a derived symmetric polynomial of
$n$-cluster form described by $(m; S_2,..., S_n)$.
From \eq{dn}, we see that $P(\{ z_i^{(a)} \})$ can also be viewed as
a symmetric polynomial of
$\ka n$-cluster form where $\ka$ is a positive integer.
When viewed as a $\ka n$-cluster polynomial,
$P(\{ z_i^{(a)} \})$ is described by
$(\ka m, S_2,..., S_{\ka n})$, where $S_{n+1},\cdots, S_{\ka n}$
are obtained from $(m; S_2,..., S_n)$ through \eq{Sbany}.

The the filling fraction $\nu=1/q$ Laughlin state
$\Phi_{1/q}=\prod_{i<j}(z_i-z_j)^q$ has a $1$-cluster form.  Thus
$\Phi_{1/q}$ can also be viewed as an $n$-cluster polynomial for
any positive $n$.  When viewed as an $n$-cluster polynomial, the
$\nu=1/q$ Laughlin state is described by
\begin{equation*}
 (m;S_2,\cdots,S_n)=
(nq; q,\cdots,\frac{qn(n-1)}{2}) .
\end{equation*}
Such a $\nu=1/q$ Laughlin state always appears as a solution of
\eq{mSn} and \eq{SanyCnd} for any $n$.

\subsection{Products of symmetric polynomials}

Let $P(\{ z_i^{(a)} \})$ and $P'(\{ z_i^{(a)} \})$ be two derived
symmetric polynomials of $n$-cluster form described by $(m; S_2,...,
S_n)$ and $(m'; S'_2,..., S'_n)$ respectively.  Then their product $
\t P(\{ z_i^{(a)} \})= P(\{ z_i^{(a)} \})P'(\{ z_i^{(a)} \})$ is
also a symmetric polynomial of $n$-cluster form.  $ \t P(\{
z_i^{(a)} \})$ is described by
\begin{equation*}
(\t m,\t S_2,\cdots,\t S_n)= (m+m'; S_2+S'_2,\cdots, S_n+S'_n)  .
\end{equation*}
This is because the pattern of zeros of
$ \t P(\{ z_i^{(a)} )$ is related to the patterns of zeros of $P(\{
z_i^{(a)}\})$ and $P'(\{ z_i^{(a)}\})$ through
\begin{equation*}
 \t D_{ab}= D_{ab} +D'_{ab}
\end{equation*}
Also the relation between $(m; S_2,..., S_n)$ and $D_{ab}$ is
linear (see \eq{DabSa}, \eq{SaDa1} and \eq{Sbany}). Therefore, if
two $\v S$-vectors, $\v S$ and $\v S'$, describe two existing
symmetric polynomials, then their sum $\t{\v S} = \v S+\v S'$ also
describes an existing symmetric polynomial, whose fillings
fractions are reciprocally additive.

Indeed, the solutions of \eq{mSn} and \eq{SanyCnd} have a structure
that is consistent with the above result. We note that \eq{mSn} and
\eq{SanyCnd} are linear in the $\v S$-vector $\v
S=(m;S_2,\cdots,S_n)$.  Thus if $\v S_1$ and $\v S_2$ are two
solutions of \eq{mSn} and \eq{SanyCnd}, then
\begin{equation}
\label{SS1S2}
\v S=k_1 \v S_1 + k_2\v S_2
\end{equation}
is also a solution for any non-negative integers $k_1$ and $k_2$.
Therefore we can divide the solutions of \eq{mSn} and \eq{SanyCnd}
into two classes: primitive solutions and non-primitive solutions.
The primitive solutions are those that cannot be written as a sum
of two other solutions.  All solutions of \eq{mSn} and
\eq{SanyCnd} are linear combinations of primitive solutions with
non-negative integral coefficients.

As an application of the product rule, let us consider a  symmetric polynomial
of $n$-cluster form $\Phi(\{z_i\})$ which is described by $(m;S_2,\cdots,S_n)$.
We can construct a new symmetric polynomial of $n$-cluster form from $\Phi(\{
z_i \})$:
\begin{equation*}
\t \Phi(\{ z_i \}) =
\Phi(\{ z_i \}) \prod_{i<j} (z_i-z_j)^q
\end{equation*}
where $q$ is even.
The symmetric polynomial $\t\Phi(\{z_i\})$
is described by
\begin{equation*}
 (\t m;\t S_2,\cdots,\t S_n)=
(m+nq; S_2+q,\cdots,S_n+\frac{qn(n-1)}{2})
\end{equation*}

\section{Some examples}

In this section, we will give some examples of symmetric polynomials
described by the $\v S$-vector $(m;S_2,\cdots,S_n)$ that satisfy
\eq{mSn}, \eq{SanyCnd}, and \eq{D3even}.

\subsection{$n=1$ cases}

If $n=1$, the different patterns of zeros are characterized by an
even integer $m$.  We find $S_a=ma(a-1)/2$ and $D_{ab}=mab$.
Each even $m$ corresponds to a $\nu=1/m$ Laughlin state
\begin{equation*}
 \Phi_{1/m}(\{z_i\})=\prod_{i<j}(z_i-z_j)^m  .
\end{equation*}

We have introduced three equivalent ways to describe a pattern of
zeros: the $\v S$-vector $(m;S_2,\cdots,S_n)$, the $\v h$-vector
$(\frac{m}{n};h^\text{sc}_1,\cdots,h^\text{sc}_n)$, and the boson
occupation number $n_l$: $(n_0,\cdots,n_{m-1})$. For the $\nu=1/m$
Laughlin state those data are given by
\begin{align*}
\Phi_{1/m}: \ \ \ \ \ \ \ \ \v S&=(m; ),
\nonumber\\
(\frac{m}{n}; h^\text{sc}_1)&=(m; 0),
\nonumber\\
(n_0,\cdots,n_{m-1})&=(1,0, \cdots,0) .
\end{align*}

\subsection{$n=2$ cases}
\label{n2case}

If $n=2$, the different patterns of zeros are characterized by two integers
$m,S_2$. The following two sets of $m,S_2$
are the primitive solutions of
\eq{mSn} and \eq{SanyCnd}:
\begin{equation*}
 (m;S_2) =(1;0),\ \ \ \ \ \
 (m';S'_2) =(4;2)
\end{equation*}

Let us discuss the solution $(m;S_2) =(1;0)$ in more detail.
The corresponding boson occupation numbers are
\begin{equation*}
 (n_0,n_1,\cdots)=(2,2,2,\cdots)
\end{equation*}
where there are two bosons occupying each orbital. Let us check
condition \eq{PJtot} for the $J=1/2$ case where there are only two
orbitals. This leads to a state $|\{2,2\}\>$ with 4 bosons
described by the wave function
\begin{equation*}
 \Phi_{\{2,2\}}=
z_1z_2+
z_1z_3+
z_1z_4+
z_2z_3+
z_2z_4+
z_3z_4
\end{equation*}
On sphere the above wave function becomes (see Section
\ref{EleSph})
\begin{equation*}
 \Phi^\text{sp}_{\{2,2\}}=\cS[v_1 v_2 u_3 u_4] .
\end{equation*}
where $\cS$ is the symmetrization operator.
Since $S_4=2$, we find that $J_4=4J-S_4=0$ and $P_{4,J_4}$ is a projection
into the subspace with vanishing total angular momentum.  A direct calculation
reveals that $P_{4,J_4}\Phi^\text{sp}_{\{2,2\}}=0$.  Thus $(m;S_2) =(1;0)$
does not satisfy the condition \eq{PJtot}, and does not correspond to any
translation invariant symmetric polynomial.

Now let consider $m,S_2$ that satisfy a new condition
\eq{D3even} in addition to \eq{mSn}, \eq{SanyCnd}.  The following two sets of
$m,S_2$ are the primitive solutions:
\begin{align*}
\Phi_{\frac{2}{2};Z_2}:\ \ \ \ (m;S_2) &=(2;0),
\nonumber\\
(\frac{m}{n}; h^\text{sc}_1,\cdots,h^\text{sc}_n) &=(\frac{2}{2}; \frac12,0),
\nonumber\\
(n_0,\cdots,n_{m-1})&=(2,0)
\end{align*}
and
\begin{align*}
\Phi_{1/2}:\ \ \ \ \ (m;S_2) &=(4;2)
\nonumber\\
(\frac{m}{n}; h^\text{sc}_1,\cdots,h^\text{sc}_n) &=(\frac{4}{2}; 0,0),
\nonumber\\
(n_0,\cdots,n_{m-1})&=(1,0,1,0)
\end{align*}
Here we also listed the corresponding $\v h$-vector $\v h=(\frac{m}{n};
h^\text{sc}_1,\cdots,h^\text{sc}_n)$ and the boson occupation numbers
$(n_0,\cdots,n_{m-1})$.

Let us discuss the solution $(m;S_2) =(2;0)$ in more detail.
We find $(S_1,S_2,S_3,S_4)=(0,0,2,4)$ and
\begin{equation*}
\bpm
D_{11}& D_{12}\\
D_{21}& D_{22}\\
\epm
=
\bpm
 0 & 2\\
 2&  4\\
\epm ,
\end{equation*}
which means that we will have no zero if we bring two particles
together and a second order zero if we bring a third particle to a
two-particle cluster. Such a pattern of zeros describes the
following translation invariant symmetric polynomial
\begin{equation}
\label{PhiPf}
\Phi_{\frac{2}{2};Z_2}(\{z_i\})= \cA\Big( \frac{1}{z_1-z_2}  \frac{1}{z_3-z_4}  \cdots \Big)
\prod_{i<j}(z_i-z_j)
\end{equation}
which is the filling-fraction $\nu=1$ bosonic Pfaffian
state.\cite{MR9162} Here $\cA$ is the antisymmetrization operator.
The  Pfaffian state can be written as a correlation of the following
operator in a CFT
\begin{equation*}
 V_e(z)=\psi(z) e^{i\phi(z)},
\end{equation*}
where $\psi(z)$ is the Majorana fermion operator in the Ising CFT (which is
also the $Z_2$ parafermion CFT).  The $h^\text{sc}_1=1/2$ in the $\v h$-vector
is the scaling dimension of $\psi$.

The other solution $(m;S_2) =(4;2)$ gives rise to the following $D_{ab}$:
\begin{equation*}
\bpm
D_{11}& D_{12}\\
D_{21}& D_{22}\\
\epm
=
\bpm
 2 & 4\\
 4&  8\\
\epm
\end{equation*}
It describes the symmetric polynomial
\begin{equation}
\label{Phi12}
\Phi_{1/2}(\{z_i\})=
\prod_{i<j}(z_i-z_j)^2
\end{equation}
which is the filling-fraction $\nu=1/2$ bosonic Laughlin state.
The  $\nu=1/2$ Laughlin state can be written as a correlation of the
following operator in the Gaussian model (or $U(1)$ CFT)
\begin{equation*}
 V_e(z)=e^{i\sqrt{2}\phi(z)} .
\end{equation*}

We note that the $\nu=1/2$ bosonic Laughlin state is characterized by a
pattern of boson occupation numbers $(n_l)=(1,0,1,0,1,0,\cdots)$, and the
$\nu=1$ bosonic Pfaffian state characterized by $(n_l)=(2,0,2,0,2,0,\cdots)$.
Those patterns match the boson occupation distributions of the two states in
the thin cylinder limit.\cite{SL0604,BKW0608,BH0737,BH0762,SY0802} This
appears to be a general result: 
\emph{the $n_l$ that characterize a symmetric polynomial correspond to one of
the boson occupation distribution of the same state in the thin cylinder
limit.}
Or more precisely:
\emph{the $n_l$ that characterize a symmetric polynomial correspond to 
the boson occupation distribution of the same state in the thin sphere
limit.}\cite{WWqp}

The solution $(m;S_2) =(4;0)=2\times (2;0)$
gives rise to the following $D_{ab}$:
\begin{equation}
\label{DabPf2}
\bpm D_{11}& D_{12}\\
     D_{21}& D_{22}\\ \epm =
\bpm 0 & 4\\
     4 & 8\\ \epm
\end{equation}
It describes a symmetric polynomial which is the square of $\Phi_{\frac{2}{2};Z_2}$
\begin{align}
\label{PhiPf2}
 \Phi_{Z_2Z_2}(\{z_i\}) &= \Phi_{\frac{2}{2};Z_2}^2(\{z_i\})
\\
&=
\left(\cA\Big( \frac{1}{z_1-z_2}  \frac{1}{z_3-z_4}  \cdots \Big) \right)^2
\prod_{i<j}(z_i-z_j)^2
\nonumber
\end{align}

Let us consider another translation invariant symmetric polynomial
\begin{equation}
\label{Phidw}
\Phi_{dw}(\{z_i\})= \cS\Big( \frac{1}{(z_1-z_2)^2}  \frac{1}{(z_3-z_4)^2}
\cdots \Big)
\prod_{i<j}(z_i-z_j)^2.
\end{equation}
where $\cS$ is the symmetrization operator. We note that
$\Phi_{dw}(\{z_i\})$ and $\Phi_{Z_2Z_2}(\{z_i\})$ has the same
pattern of zeros given by \eq{DabPf2}. In the following, we would
like to show that
\begin{equation*}
 \Phi_{dw}(\{z_i\})\propto  \Phi_{Z_2Z_2}(\{z_i\}) .
\end{equation*}

We first note that
$\Phi_{Z_2Z_2}(\{z_i\})$ can be written as a correlation of the
following operator in a CFT
\begin{equation*}
 V_e(z)=\la_1(z)\la_2(z) e^{i\phi(z)}
\end{equation*}
where $\la_1(z)$ is the Majorana fermion operator in a Ising CFT
and $\la_2(z)$ is the Majorana fermion operator in another Ising CFT.
Thus the operator $V_e(z)=\la_1(z)\la_2(z) e^{i\phi(z)}$ is an operator
in Ising$\times$Ising$\times U(1)$ CFT.

The state $\Phi_{dw}(\{z_i\})$ can also be written as a correlation of the
following operator in a CFT
\begin{equation*}
 V_e(z)=\prt_z \t\phi(z) e^{i\phi(z)}
\end{equation*}
where $\t\phi(z)$ is the field in a second $U(1)$ CFT. Thus the
operator $V_e(z)=\prt_z\phi(z) e^{i\phi(z)}$ is an operator in
$U(1)\times \t U(1)$ CFT.

From the bosonization of the Ising$\times$Ising CFT, one can show that the
Ising$\times$Ising
CFT is equivalent to the $\t U(1)$ CFT and $\la_1\la_2$ has the same $N$-body
correlation functions as $\prt_z \t\phi(z)$. Thus
$\Phi_{dw}(\{z_i\})=\Phi_{Z_2Z_2}(\{z_i\})$.
Numerical calculations have suggested that
$\Phi_{dw}(\{z_i\})$ has gapless excitation and is unstable.

Next, let us consider the following two polynomials
\begin{align}
\label{PhiPfq}
 \Phi_{(Z_2)^q}(\{z_i\}) &= \Phi_{\frac{2}{2};Z_2}^q(\{z_i\})
\\
&=
\left(\cA\Big( \frac{1}{z_1-z_2}  \frac{1}{z_3-z_4}  \cdots \Big) \right)^q
\prod_{i<j}(z_i-z_j)^q
\nonumber
\end{align}
and
\begin{equation}
\label{Phiqw} \Phi_{qw}(\{z_i\})= \cS_q\Big( \frac{1}{(z_1-z_2)^q}
\frac{1}{(z_3-z_4)^q} \cdots \Big) \prod_{i<j}(z_i-z_j)^q,
\end{equation}
where $\cS_q=\cS$ when $q=$ even and $\cS_q=\cA$ when $q=$ odd.  The two
symmetric polynomials have the same pattern of zeros $D_{ab}$.  But when
$q>2$, the two polynomials are different.  Those polynomials
provide us examples that there can be more than one polynomials that have the
same pattern of zeros.

\subsection{$n=3$ cases}

If $n=3$, the different patterns of zeros are characterized by three integers
$m,S_2,S_3$. The following two sets of $m,S_2,S_3$ are the primitive
solutions of \eq{mSn} and \eq{SanyCnd}:
\begin{align*}
\Phi_{\frac{3}{2};Z_3}:\ \ \ \ (m;S_2,S_3) &=(2;0,0),
\nonumber\\
(\frac{m}{n}; h^\text{sc}_1,\cdots,h^\text{sc}_n)
  &=(\frac{2}{3}; \frac23,\frac23,0),
\nonumber\\
(n_0,\cdots,n_{m-1})&=(3,0)
\end{align*}
and
\begin{align*}
\Phi_{1/2}:\ \ \ \ \ (m;S_2,S_3) &=(6;2,6)
\nonumber\\
(\frac{m}{n}; h^\text{sc}_1,\cdots,h^\text{sc}_n) &=(\frac{6}{3}; 0,0,0),
\nonumber\\
(n_0,\cdots,n_{m-1})&=(1,0,1,0,1,0)
\end{align*}
When $n=$ odd, we find that the solutions of \eq{mSn} and \eq{SanyCnd}
automatically satisfy \eq{D3even}.

From the $\v h$-vector of the solution $(m;S_2,S_3) =(2;0,0)$, we find that
the corresponding polynomial $\Phi_{\frac{3}{2};Z_3}$ describes the $Z_3$ Read-Rezayi
parafermion state,\cite{RR9984} since $h^\text{sc}_1=h^\text{sc}_2=2/3$ in the
$\v h$-vector match the scaling dimensions of the simple-current operators in
the $Z_3$ parafermion CFT.  Such a state has a filling-fraction $\nu=3/2$.
Here we have been using
$\Phi_{\frac{n}{m};Z_n}$ to denote a $Z_n$ parafermion state.
We will follow such a convention in the rest of this paper.
The second solution $(m;S_2,S_3) =(6;2,6)$ describes the $\nu=1/2$ Laughlin
state.

\subsection{$n=4$ cases}

When $n=4$, the different patterns of zeros are characterized by
four integers $m,S_2,S_3,S_4$.  The primitive solutions of \eq{mSn}
and \eq{SanyCnd} are given by the following three sets of
$m,S_2,S_3,S_4$:
\begin{equation*}
(m;S_2,S_3,S_4)=(1;0,0,0),\ (2;0,1,2),\ (8,2,6,12) .
\end{equation*}

The solution $\v S= (2;0,1,2)$ is the same as solution $\v S=(1;0)$ for the
$n=2$ case (\ie the two solutions give rise to the same sequence $\{S_a\}$,
$a=1,2,3,\cdots$). Such a solution does not satisfy \eq{PJtot} as shown in
section \ref{n2case}.

The solution $\v S= (1;0,0,0)$ does not satisfy \eq{PJtot} either.
Let us check the condition \eq{PJtot} for $J=1/2$ case where there
are only two orbitals. This leads to a state $|\{4,4\}\>$ with 8
bosons. On sphere, such a state is given by
\begin{equation*}
 \Phi^\text{sp}_{\{4,4\}}=\cS[v_1v_2v_3v_4u_5u_6u_7u_8] .
\end{equation*}
Since $S_8=4$, we find that $J_8=8J-S_8=0$ and $P_{8,J_8}$ is a projection
into the subspace with vanishing total angular momentum.  Explicit calculation
shows that the state $\Phi^\text{sp}_{\{4,4\}}$ has a vanishing projection
onto the $J_\text{tot}=0$ subspace.

So we consider the solutions of \eq{mSn}, \eq{SanyCnd}, and \eq{D3even} to
exclude those invalid cases.  The primitive solutions of \eq{mSn}, \eq{SanyCnd},
and \eq{D3even} are
\begin{align}
\label{n4Z4}
\Phi_{\frac{4}{2};Z_4}: (m;S_2,\cdots,S_n) &=(2;0,0,0),
\nonumber\\
(\frac{m}{n}; h^\text{sc}_1,\cdots,h^\text{sc}_n)
  &=(\frac{2}{4}; \frac34,1,\frac34,0),
\nonumber\\
(n_0,\cdots,n_{m-1})&=(4,0),
\end{align}
\begin{align}
\label{n4Pf}
\Phi_{\frac{2}{2};Z_2}: (m;S_2,\cdots,S_n) &=(4;0,2,4),
\nonumber\\
(\frac{m}{n}; h^\text{sc}_1,\cdots,h^\text{sc}_n)
  &=(\frac{4}{4}; \frac12,0,\frac12,0),
\nonumber\\
(n_0,\cdots,n_{m-1})&=(2,0,2,0),
\end{align}
\begin{align*}
\Phi_{1/2}: (m;S_2,\cdots,S_n) &=(8;2,6,12)
\nonumber\\
(\frac{m}{n}; h^\text{sc}_1,\cdots,h^\text{sc}_n) &=(\frac{8}{4}; 0,0,0,0),
\nonumber\\
(n_0,\cdots,n_{m-1})&=(1,0,1,0,1,0,1,0) .
\end{align*}

Among those three primitive solutions, only $\Phi_{\frac{4}{2};Z_4}$ is new.
From  the $\v h$-vector in \eq{n4Z4}, we find that the solution
$(m;S_2,S_3,S_4)=(2;0,0,0)$ describes the $Z_4$ parafermion state $\Phi_{\frac{4}{2};Z_4}$
with $\nu=2$.

The solution $(m;S_2,S_3,S_4)=(4;0,2,4)$ is the same as the $\nu=1$
bosonic Pfaffian state $(m;S_2)=(2;0)$ discussed before.  So the $\v
h$-vectors of the two solutions, $(\frac{m}{n};
h^\text{sc}_1,\cdots,h^\text{sc}_4)=(\frac{4}{4};
\frac12,0,\frac12,0)$ and $(\frac{m}{n};
h^\text{sc}_1,h^\text{sc}_2)=(\frac{2}{2}; \frac12,0)$, characterize
the same state.  In fact, the repeated $(\frac12,0)$ pattern in
$(h^\text{sc}_1,\cdots,h^\text{sc}_4)$ implies that $(\frac{m}{n};
h^\text{sc}_1,\cdots,h^\text{sc}_4)=(\frac{4}{4};
\frac12,0,\frac12,0)$ can be reduced to $(\frac{m}{n};
h^\text{sc}_1,h^\text{sc}_2)=(\frac{2}{2}; \frac12,0)$.  Also the
two solutions lead to the same pattern of boson occupation numbers:
$(n_l)=(2,0,2,0,2,0,2,0,\cdots)$, again indicating that the two
solutions describe the same state.

The solution $(m;S_2,S_3,S_4)=(8;2,6,12)$ describes the $\nu=1/2$ Laughlin
state $\Phi_{1/2}$ characterized by
$(\frac{m}{n}; h^\text{sc}_1,\cdots,h^\text{sc}_4)=(\frac{8}{4}; 0,0,0,0)$.
We have seen that
$\v h_{1/2}=(\frac{2}{1}; 0)$,
$\v h_{2/4}=(\frac{4}{2}; 0,0)$,
$\v h_{3/6}=(\frac{6}{3}; 0,0,0)$, and
$\v h_{4/8}=(\frac{8}{4}; 0,0,0,0)$
all describe the same $\nu=1/2$ Laughlin state $\Phi_{1/2}$.
All the above solutions share the same pattern of boson
occupation numbers: $(n_l)=(1,0,1,0,1,0,1,0,\cdots)$

The product state $\Phi_{\frac{4}{2};Z_4}\Phi_{\frac{2}{2};Z_2}$ described by
\begin{align*}
\Phi_{\frac{4}{2};Z_4}\Phi_{\frac{2}{2};Z_2}: (m;S_2,\cdots,S_n) &=(6;0,2,4)
\nonumber\\
(\frac{m}{n}; h^\text{sc}_1,\cdots,h^\text{sc}_n)
  &=(\frac{6}{4}; \frac54,1,\frac54,0),
\nonumber\\
(n_0,\cdots,n_{m-1})&=(2,0,2,0,0,0) .
\end{align*}
is a possible stable $\nu=2/3$ FQH state.  Note that
the above $\v h$-vector is the sum of the $\v h$-vectors of the $Z_2$ and
$Z_4$ parafermion states.

\subsection{$n=5$ cases}

When $n=5$,  the conditions \eq{mSn} and \eq{SanyCnd} have the following three
sets of primitive solutions:
\begin{align*}
\Phi_{\frac{5}{2};Z_5}: (m;S_2,\cdots,S_n) &= (2;0,0,0,0),
\nonumber\\
(\frac{m}{n}; h^\text{sc}_1,\cdots,h^\text{sc}_n) &=
(\frac{2}{5}; \frac45,\frac65,\frac65,\frac45,0) ;
\nonumber\\
(n_0,\cdots,n_{m-1})&=(5,0) .
\end{align*}
\begin{align*}
\Phi_{\frac{5}{8};Z_5^{(2)}}: (m;S_2,\cdots,S_n) &= (8;0,2,6,10),
\nonumber\\
(\frac{m}{n}; h^\text{sc}_1,\cdots,h^\text{sc}_n) &=
(\frac{8}{5}; \frac65,\frac45,\frac45,\frac65,0) ;
\nonumber\\
(n_0,\cdots,n_{m-1})&=(2,0,1,0,2,0,0,0) .
\end{align*}
\begin{align*}
\Phi_{1/2}:
(m;S_2,\cdots,S_n) &= (10;2,6,12,20) ,
\nonumber\\
(\frac{m}{n}; h^\text{sc}_1,\cdots,h^\text{sc}_n) &=
(\frac{10}{5}; 0,0,0,0,0) .
\nonumber\\
(n_0,\cdots,n_{m-1})&=(1,0,1,0,1,0,1,0,1,0) .
\end{align*}
All other solutions
are linear combinations of the above three solutions.

$(\frac{m}{n}; h^\text{sc}_1,\cdots,h^\text{sc}_5) =
(\frac{2}{5}; \frac45,\frac65,\frac65,\frac45,0)$
describes a $Z_5$ parafermion state
$\Phi_{\frac{5}{2};Z_5}$ studied by Read and Rezayi.\cite{RR9984}
$(\frac{m}{n}; h^\text{sc}_1,\cdots,h^\text{sc}_5) =
(\frac{8}{5}; \frac65,\frac45,\frac45,\frac65,0)$
describes a new
parafermion state $\Phi_{\frac{5}{8};Z^{(2)}_5}$ with $\nu=5/8$.
The third state
$(\frac{m}{n}; h^\text{sc}_1,\cdots,h^\text{sc}_5) =
(\frac{10}{5}; 0,0,0,0,0)$
describes the $\nu=1/2$ Laughlin state
$\Phi_{1/2}$.

The $Z_5$ parafermion state $\Phi_{\frac{5}{2};Z_5}$ can be expressed as a
correlation of simple-current operators $\psi_1$ in the  $Z_5$
parafermion CFT.  $\psi_1$ has a scaling dimension of
$h^\text{sc}_1=4/5$.  The new parafermion state $\Phi_{\frac{5}{8};Z^{(2)}_5}$
can be expressed as a correlation of simple-current operators
$\psi_2$ in the  $Z_5$ parafermion CFT.  $\psi_2$ has a scaling
dimension of $h^\text{sc}_2=6/5$. In general, the simple-current
operator $\phi_l$ of a $Z_n$ parafermion CFT has a scaling dimension
\begin{equation*}
h^\text{sc}_l=\frac{l(n-l)}{n}.
\end{equation*}
Here we have been using $\Phi_{\frac{n}{m};Z_n^{(k)}}$ to denote a generalized
$Z_n$ parafermion state.  We will follow such a convention in the rest of this
paper.

\subsection{$n=6$ cases}

When $n=6$,
the conditions \eq{mSn}, \eq{SanyCnd}, and \eq{D3even} have the
following four sets of primitive solutions:
\begin{align*}
\Phi_{\frac{6}{2};Z_6}: (m;S_2,\cdots,S_n) &= (2;0,0,0,0,0),
\nonumber\\
(\frac{m}{n}; h^\text{sc}_1,\cdots,h^\text{sc}_n) &=
(\frac{2}{6}; \frac56,\frac43,\frac32,\frac43,\frac56,0) ;
\nonumber\\
(n_0,\cdots,n_{m-1})&=(6,0) .
\end{align*}
$\Phi_{\frac{2}{2};Z_2}$, $\Phi_{\frac{3}{2};Z_3}$, and $\Phi_{1/2}$.  Three of the four primitive
solutions have been discussed before and only one solution, $\Phi_{\frac{6}{2};Z_6}$, is
new.  The $\Phi_{\frac{6}{2};Z_6}$ state is the $Z_6$ parafermion state.\cite{RR9984}
$\Phi_{\frac{2}{2};Z_2}$ and $\Phi_{\frac{3}{2};Z_3}$ are the $Z_2$ and $Z_3$ parafermion states
discussed before.

Using $\Phi_{\frac{2}{2};Z_2}$, $\Phi_{\frac{3}{2};Z_3}$, and $\Phi_{\frac{6}{2};Z_6}$, we can construct
some interesting and possibly stable composite states:
\begin{align*}
\Phi_{\frac{3}{2};Z_3}\Phi_{\frac{6}{2};Z_6}: (m;S_2,\cdots,S_n) &= (6;0,0,2,4,6),
\nonumber\\
(\frac{m}{n}; h^\text{sc}_1,\cdots,h^\text{sc}_n) &=
(\frac{6}{6}; \frac32,2,\frac32,2,\frac32,0) ,
\nonumber\\
(n_0,\cdots,n_{m-1})&=(3,0,3,0,0,0) ;
\end{align*}
\begin{align*}
\Phi_{\frac{2}{2};Z_2}\Phi_{\frac{6}{2};Z_6}: (m;S_2,\cdots,S_n) &= (8;0,2,4,8,12),
\nonumber\\
(\frac{m}{n}; h^\text{sc}_1,\cdots,h^\text{sc}_n) &=
(\frac{8}{6}; \frac43,\frac43,2,\frac43,\frac43,0) ,
\nonumber\\
(n_0,\cdots,n_{m-1})&=(2,0,2,0,2,0,0,0) ;
\end{align*}
\begin{align*}
\Phi_{\frac{2}{2};Z_2}\Phi_{\frac{3}{2};Z_3}: (m;S_2,\cdots,S_n) &= (10;0,2,6,12,18),
\nonumber\\
(\frac{m}{n}; h^\text{sc}_1,\cdots,h^\text{sc}_n) &=
(\frac{10}{6}; \frac76,\frac23,\frac12,\frac23,\frac76,0) ,
\nonumber\\
(n_0,\cdots,n_{m-1})&=(2,0,1,0,1,0,2,0,0,0) ;
\end{align*}
\begin{align*}
\Phi_{\frac{2}{2};Z_2}\Phi_{\frac{3}{2};Z_3}\Phi_{\frac{6}{2};Z_6}: (m;S_2,\cdots,S_n) &= (12;0,2,6,12,18),
\nonumber\\
(\frac{m}{n}; h^\text{sc}_1,\cdots,h^\text{sc}_n) &=
(\frac{12}{6}; 2,2,2,2,2,0) ,
\nonumber\\
(n_0,\cdots,n_{m-1})=(2,0,1,&\,0,1,0,2,0,0,0,0,0) .
\end{align*}
The filling fractions of those states are given by $\nu=n/m$.

\subsection{$n=7$ cases}

When $n=7$,  the conditions \eq{mSn} and \eq{SanyCnd} have the following five
sets of primitive solutions:
\begin{align*}
\Phi_{\frac{7}{2};Z_7}: (m;S_2,\cdots,S_n) &= (2;0,0,0,0,0,0),
\nonumber\\
(\frac{m}{n}; h^\text{sc}_1,\cdots,h^\text{sc}_n) &=
(\frac{2}{7}; \frac67,\frac{10}{7},\frac{12}{7},\frac{12}{7},\frac{10}{7},\frac67,0) ,
\nonumber\\
(n_0,\cdots,n_{m-1})&=(7,0) ;
\end{align*}
\begin{align*}
\Phi_{\frac{7}{8};Z^{(2)}_7}: (m;S_2,\cdots,S_n) &= (8;0,0,2,6,10,14),
\nonumber\\
(\frac{m}{n}; h^\text{sc}_1,\cdots,h^\text{sc}_n) &=
(\frac{8}{7}; \frac{10}{7},\frac{12}{7},\frac67,\frac67,\frac{12}{7},\frac{10}{7},0) ,
\nonumber\\
(n_0,\cdots,n_{m-1})&=(3,0,1,0,3,0,0,0) ;
\end{align*}
\begin{align*}
\Phi_{\frac{7}{18};Z^{(3)}_7}: (m;S_2,\cdots,S_n) &= (18;0,4,10,18,30,42),
\nonumber\\
(\frac{m}{n}; h^\text{sc}_1,\cdots,h^\text{sc}_n) &=
(\frac{18}{7}; \frac{12}{7},\frac67,\frac{10}{7},\frac{10}{7},\frac67,\frac{12}{7},0) ,
\nonumber\\
(n_0,\cdots,n_{m-1})=(2,&\,0,0,0,0,1,0,0,0,2,0,0,0,0,0) ;
\end{align*}
\begin{align}
\label{Phiz17}
\Phi_{\frac{7}{14};C_7}: (m;S_2,\cdots,S_n) &= (14;0,2,6,12,20,28),
\\
(\frac{m}{n}; h^\text{sc}_1,\cdots,h^\text{sc}_n) &=
(\frac{14}{7}; 2,2,2,2,2,2,0) ,
\nonumber\\
(n_0,\cdots,n_{m-1})=(2,&\,0,1,0,1,0,1,0,2,0,0,0,0,0) ;
\nonumber
\end{align}
and $\Phi_{1/2}$.

$\Phi_{\frac{7}{2};Z_7}$ is the $Z_7$ parafermion state  which can be
expressed as a correlation of simple-current operators $\psi_1$ in
the  $Z_7$ parafermion CFT.  $\psi_1$ has a scaling dimension of
$h^\text{sc}_1=6/7$. $\Phi_{\frac{7}{8};Z^{(2)}_7}$ and $\Phi_{\frac{7}{18};Z^{(3)}_7}$ are
two new $Z_7$ parafermion states. $\Phi_{\frac{7}{8};Z^{(2)}_7}$ can be
expressed as a correlation of simple-current operators $\psi_2$
while $\Phi_{\frac{7}{18};Z^{(3)}_7}$ can be expressed as a correlation of
simple-current operators $\psi_3$ in the  $Z_7$ parafermion CFT.
$\psi_2$ has a scaling dimension of $h^\text{sc}_2=10/7$ and
$\psi_3$ has a scaling dimension of $h^\text{sc}_3=12/7$.

Let us discuss the state $\Phi_{\frac{7}{14};C_7}$ in more detail.
$\Phi_{\frac{7}{14};C_7}$ has a form
\begin{equation*}
 \Phi_{\frac{7}{14};C_7}(\{z_i\}) = G_{C_7}(\{z_i\}) \prod(z_i-z_j)^2
\end{equation*}
where the pattern of the zeros (or poles) for $G_{C_7}(\{z_i\})$ is
given by (see \eq{Dabdab})
\begin{equation*}
 (d_{ab})=
\bpm
 -2&-2&-2&-2&-2&-4&0 \\
 -2&-2&-2&-2&-4&-2&0 \\
 -2&-2&-2&-4&-2&-2&0 \\
 -2&-2&-4&-2&-2&-2&0 \\
 -2&-4&-2&-2&-2&-2&0 \\
 -4&-2&-2&-2&-2&-2&0 \\
  0& 0& 0& 0& 0& 0&0 \\
\epm
\end{equation*}
where $a,b=1,\cdots,7$. It implies that there is a second order
pole as an $a$-cluster approaches a $b$-cluster if $a+b<7$ and a
fourth order pole as an $a$-cluster approaches a $b$-cluster if
$a+b=7$. Such a pattern of poles is reproduced by
\begin{align*}
 G_{C_7}(\{z_i\})=\cS\Big[
f^2_{C_7}(z_1,\cdots,z_7)
f^2_{C_7}(z_8,\cdots,z_{14})
\cdots
\Big]
\end{align*}
where
\begin{equation*}
 f_{C_n}(z_1,\cdots,z_n)=
\frac{1}{z_1-z_2}
\frac{1}{z_2-z_3}\cdots
\frac{1}{z_{n-1}-z_n}
\frac{1}{z_n-z_1} .
\end{equation*}

\begin{figure}[tb]
\centerline{
\includegraphics[scale=0.5]{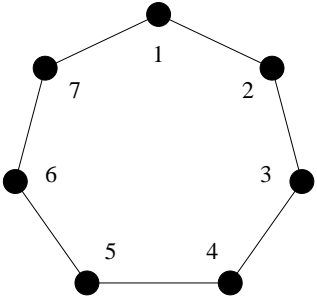}
}
\caption{
The graph that represents
$f_{C_7}$. Each line between
the $i^\text{th}$ dot and
the $j^\text{th}$ dot represent a factor
$1/(z_i-z_j)$.
}
\label{z17}
\end{figure}

To confirm such a result, we note that, for $a=2,\cdots,7$, the
minimal total powers of $a$ variables in $f^2_{C_7}$ are $s_a$ with
$(s_2,\cdots,s_7)=(-2,-4,-6,-8,-10,-14)$ (see \eq{saSa}). The
minimal total powers of $a$ variables in $\prod(z_i-z_j)^2$ are $\t
s_a$ with $(\t s_2,\cdots,\t s_7)=(2,6,12,20,30,42)$. Thus the
minimal total powers of $a$ variables in $\Phi_{\frac{7}{14};C_7}$ are given by
$S_a=s_a+\t s_a$: $(S_2,\cdots,S_7)=(0,2,6,12,20,28)$ which agrees
with \eq{Phiz17}.

The function $f_{C_7}(z_1,\cdots,z_7)$
can be represented graphically as in Fig. \ref{z17}.
In such a  graphic representation,
the maximum total order of the poles of $a$ variables
is the maximum number of lines that connect $a$ dots.
Note that the maximum total order of the poles of $a$ variables
is the negative of the minimal total power of zeros
of $a$ variables.

In fact, for any $n$, we have a state $\Phi_{\frac{n}{2n};C_n}$ described by
$(h^\text{sc}_1,\cdots, h^\text{sc}_{n-1}, h^\text{sc}_n)=(2,\cdots,2,0)$.
The explicit wave function is given by
\begin{align*}
\Phi_{\frac{n}{2n};C_n} &= \prod_{i<j} (z_i-z_j)^2 \times
\nonumber\\
&\ \ \ \cS\Big[
f^2_{C_n}(z_1,\cdots,z_n)
f^2_{C_n}(z_{n+1},\cdots,z_{2n})
\cdots
\Big]
\end{align*}

\subsection{$n=8$ cases}

\begin{figure}[tb]
\centerline{
\includegraphics[scale=0.5]{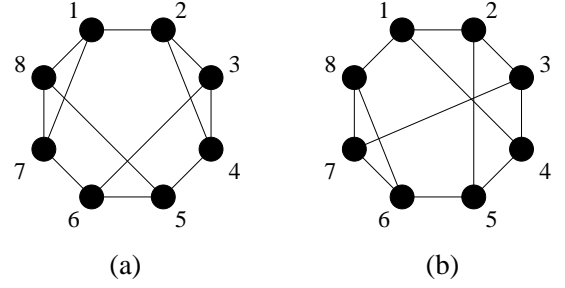}
}
\caption{
The graphs that represent $f_{C_8/Z_2}$. Each line between the $i^\text{th}$
dot and the $j^\text{th}$ dot represent a factor $1/(z_i-z_j)$.  After the
antisymmetrization, $f_{C_8/Z_2}$ from graph (a) gives rise to a non-zero
antisymmetric function, while $f_{C_8/Z_2}$ from graph (b) gives rise to
vanishing antisymmetric function.
}
\label{C8Z2}
\end{figure}

When $n=8$, the conditions \eq{mSn}, \eq{SanyCnd}, and \eq{D3even} have the
following six sets of primitive solutions:
\begin{align*}
\Phi_{\frac{8}{2};Z_8}: (m;S_2,\cdots,S_n) &= (2;0,0,0,0,0,0,0),
\nonumber\\
(\frac{m}{n}; h^\text{sc}_1,\cdots,h^\text{sc}_n) &=
(\frac{2}{8}; \frac78,\frac32,\frac{15}{8},2,\frac{15}{8},\frac32,\frac78,0) ,
\nonumber\\
(n_0,\cdots,n_{m-1})&=(8,0) ;
\end{align*}
\begin{align*}
\Phi_{\frac{8}{18};Z_8^{(3)}}: (m;S_2,\cdots,S_n) &= (18;0,2,8,14,24,36,48),
\nonumber\\
(\frac{m}{n}; h^\text{sc}_1,\cdots,h^\text{sc}_n) &=
(\frac{18}{8}; \frac{15}{8},\frac32,\frac78,2,\frac78,\frac32,\frac{15}{8},0) ,
\nonumber\\
(n_0,\cdots,n_{m-1})=(2,0,1,&\,0,0,0,2,0,0,0,1,0,2,0,0,0,0,0) ;
\end{align*}
\begin{align*}
\Phi_{\frac{8}{8};C_8/Z_2}: (m;S_2,\cdots,S_n) &= (8;0,0,2,4,8,12,16),
\nonumber\\
(\frac{m}{n}; h^\text{sc}_1,\cdots,h^\text{sc}_n) &=
(\frac{8}{8}; \frac32,2,\frac32,2,\frac32,2,\frac32,0) ,
\nonumber\\
(n_0,\cdots,n_{m-1})&=(3,0,2,0,3,0,0,0) ;
\end{align*}
$\Phi_{\frac{4}{2};Z_4}$, $\Phi_{\frac{2}{2};Z_2}$, and $\Phi_{1/2}$.

$\Phi_{\frac{8}{2};Z_8}$ is the $Z_8$ parafermion state  which can be expressed
as a correlation of simple-current operators $\psi_1$ in the  $Z_8$
parafermion CFT.  $\psi_1$ has a scaling dimension of
$h^\text{sc}_1=7/8$. $\Phi_{\frac{8}{18};Z^{(3)}_8}$ is a new $Z_8$ parafermion
state.  $\Phi_{\frac{8}{18};Z^{(3)}_8}$ can be expressed as a correlation of
simple-current operators $\psi_3$ in the  $Z_8$ parafermion CFT.
$\psi_3$ has a scaling dimension of $h^\text{sc}_3=15/8$.

Let us discuss the state $\Phi_{\frac{8}{8};C_8/Z_2}$ in more detail.
We note that the $\v h$-vector for the
$\Phi_{\frac{8}{8};C_8/Z_2}$ state is the difference of the $\v h$-vectors of the
$\Phi_{\frac{8}{2};C_8}$ state and the $\Phi_{\frac{2}{2};Z_2}$ state:
\begin{align*}
&\ \ \ (\frac32,2,\frac32,2,\frac32,2,\frac32,0)
\nonumber\\
&=
(2,2,2,2,2,2,2,0)- (\frac12,0,\frac12,0,\frac12,0,\frac12,0) .
\end{align*}
$\Phi_{\frac{8}{8};C_8/Z_2}$ has a form
\begin{equation*}
 \Phi_{\frac{8}{8};C_8/Z_2}(\{z_i\}) = G_{C_8/Z_2}(\{z_i\}) \prod(z_i-z_j)
\end{equation*}
where the minimal total power of $a$ variables in
$G_{C_8/Z_2}(\{z_i\})$ is given by $s_a$ with
$(s_2,\cdots,s_8)=(-1,-3,-4,-6,-7,-9,-12)$.
We find that
\begin{align*}
 G_{C_8/Z_2}(\{z_i\})=\cA\Big[
f_{C_8/Z_2}(z_1,\cdots,z_8)
f_{C_8/Z_2}(z_9,\cdots,z_{16})
\cdots
\Big]
\end{align*}
where $\cA$ is the antisymmetrization operator and
the function $f_{C_8/Z_2}(z_1,\cdots,z_8)$
is represented by Fig. \ref{C8Z2}(a).

\begin{figure}[tb]
\centerline{
\includegraphics[scale=0.5]{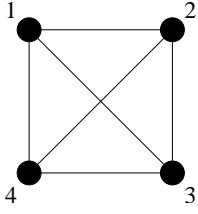}
}
\caption{
The graph that represents $f_{C_4/Z_2}$. Each line between the $i^\text{th}$
dot and the $j^\text{th}$ dot represent a factor $1/(z_i-z_j)$.
}
\label{C4Z2}
\end{figure}

We would like to mention that a state simpler than $\Phi_{\frac{8}{8};C_8/Z_2}$
is $\Phi_{\frac{4}{4};C_4/Z_2}$ that has the same pattern of zeros as a
composite state of $Z_4$ parafermion state $\Phi_{\frac{4}{2};Z_4}$:
\begin{equation*}
 \Phi_{\frac{4}{4};C_4/Z_2}\approx \Phi_{\frac{4}{2};Z_4}\Phi_{\frac{4}{2};Z_4}
\end{equation*}
Here $\approx$ means to have the same pattern of zeros.
$\Phi_{\frac{4}{4};C_4/Z_2}$ has a form
\begin{equation*}
 \Phi_{\frac{4}{4};C_4/Z_2}(\{z_i\}) = G_{C_4/Z_2}(\{z_i\}) \prod(z_i-z_j)
\end{equation*}
where the minimal total power of $a$ variables in
$G_{C_4/Z_2}(\{z_i\})$ is given by $s_a$ with
$(s_2,\cdots,s_4)=(-1,-3,-6)$.
We find that
\begin{align*}
 G_{C_4/Z_2}(\{z_i\})=\cA\Big[
f_{C_4/Z_2}(z_1,\cdots,z_4)
f_{C_4/Z_2}(z_5,\cdots,z_{8})
\cdots
\Big]
\end{align*}
where the function $f_{C_4/Z_2}(z_1,\cdots,z_4)$ is represented by Fig.
\ref{C4Z2}.  In fact, $f_{C_4/Z_2}(z_1,\cdots,z_4)=\prod_{1\leq i<j\leq
4}\frac{1}{z_i-z_j}$ is the only function whose total order of poles for $2$-,
$3$-, and $4$-particle clusters are given by $1$, $3$, and $6$ respectively.
Such a state is studied recently by Yue Yu.\cite{YueYu}

\begin{figure}[tb]
\centerline{
\includegraphics[scale=0.5]{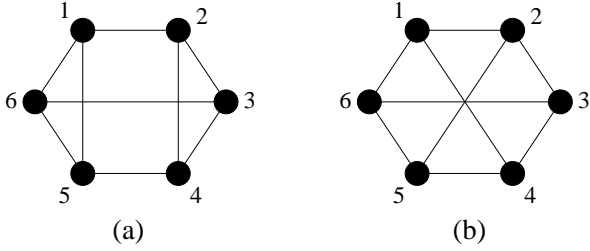}
} \caption{ The graphs that represent $f_{C_6/Z_2}$. Each line
between the $i^\text{th}$ dot and the $j^\text{th}$ dot represent a
factor $1/(z_i-z_j)$.  After the antisymmetrization, $f_{C_6/Z_2}$
from graph (a) gives rise to a non-zero antisymmetric function,
while $f_{C_6/Z_2}$ from graph (b) vanishes. } \label{C6Z2}
\end{figure}

Another interesting state is $\Phi_{\frac{6}{6};C_6/Z_2}$ which has the same pattern of
zeros as a composite state of $Z_3$ and $Z_6$ parafermion states:
\begin{equation*}
 \Phi_{\frac{6}{6};C_6/Z_2}\approx \Phi_{\frac{3}{2};Z_3}\Phi_{\frac{6}{2};Z_6}
\end{equation*}
$\Phi_{\frac{6}{6};C_6/Z_2}$ has a form
\begin{equation*}
 \Phi_{\frac{6}{6};C_6/Z_2}(\{z_i\}) = G_{C_6/Z_2}(\{z_i\}) \prod(z_i-z_j)
\end{equation*}
where the minimal total power of $a$ variables in
$G_{C_6/Z_2}(\{z_i\})$ is given by $s_a$ with
$(s_2,\cdots,s_6)=(-1,-3,-4,-6,-9)$.
We find that
\begin{align*}
 G_{C_6/Z_2}(\{z_i\})=\cA\Big[
f_{C_6/Z_2}(z_1,\cdots,z_6)
f_{C_6/Z_2}(z_7,\cdots,z_{12})
\cdots
\Big]
\end{align*}
where the function $f_{C_6/Z_2}(z_1,\cdots,z_6)$ is represented by Fig.
\ref{C6Z2}(a).

\subsection{$n=9$ cases}

When $n=9$, the conditions \eq{mSn}, \eq{SanyCnd}, and \eq{D3even} have the
following six sets of primitive solutions:
\begin{align*}
\Phi_{\frac{9}{2};Z_9}: (m;S_2,\cdots,S_n) &= (2;0,0,0,0,0,0,0,0),
\nonumber\\
(\frac{m}{n}; h^\text{sc}_1,\cdots,h^\text{sc}_n) &=
(\frac{2}{9}; \frac89,\frac{14}{9},2,\frac{20}{9},\frac{20}{9},2,\frac{14}{9},\frac89,0) ,
\nonumber\\
(n_0,\cdots,n_{m-1})&=(9,0) ;
\end{align*}
\begin{align*}
\Phi_{\frac{9}{8};Z_9^{(2)}}: (m;S_2,\cdots,S_n) &= (8;0,0,0,2,6,10,14,18),
\nonumber\\
(\frac{m}{n}; h^\text{sc}_1,\cdots,h^\text{sc}_n) &=
(\frac{8}{9}; \frac{14}{9},\frac{20}{9},2,\frac89,\frac89,2,\frac{20}{9},\frac{14}{9},0) ,
\nonumber\\
(n_0,\cdots,n_{m-1})&=(4,0,1,0,4,0,0,0) ;
\end{align*}
\begin{widetext}
\begin{align*}
\Phi_{\frac{9}{32};Z_9^{(4)}}: (m;S_2,\cdots,S_n) &=(32;0,6,14,26,42,60,84,108),
\nonumber\\
(\frac{m}{n}; h^\text{sc}_1,\cdots,h^\text{sc}_n) &=
(\frac{32}{9}; \frac{20}{9},\frac{8}{9},2,\frac{14}{9},\frac{14}{9},2,\frac{8}{9},\frac{20}{9},0) ,
\nonumber\\
(n_0,\cdots,n_{m-1})&=
(2,0,0,0,0,0,1,0,1,0,0,0,1,0,0,0,1,0,1,0,0,0,0,0,2,0,0,0,0,0,0,0) ;
\end{align*}
\end{widetext}
\begin{align*}
\Phi_{\frac{9}{12};C_9/Z_3}: (m;S_2,\cdots,S_n) &= (12;0,2,4,8,14,20,28,36),
\nonumber\\
(\frac{m}{n}; h^\text{sc}_1,\cdots,h^\text{sc}_n) &=
(\frac{12}{9}; \frac43,\frac43,2,\frac43,\frac43,2,\frac43,\frac43,0) ,
\nonumber\\
(n_0,\cdots,n_{m-1})&=(2,0,2,0,1,0,2,0,2,0,0,0) .
\end{align*}
$\Phi_{\frac{3}{2};Z_3}$, and $\Phi_{1/2}$.

$\Phi_{\frac{9}{2};Z_9}$ is the old $Z_9$ parafermion state.
$\Phi_{\frac{9}{8};Z^{(2)}_9}$ and $\Phi_{\frac{9}{32};Z^{(4)}_9}$ are new $Z_9$
parafermion states, which can be expressed as a correlation of
simple-current operators $\psi_2$ and $\psi_4$ in the  $Z_9$
parafermion CFT, respectively.  We also note that the $\v
h$-vector for the $\Phi_{\frac{9}{12};C_9/Z_3}$ state is the difference of the
$\v h$-vectors of the $\Phi_{\frac{9}{18};C_9}$ state and the $\Phi_{\frac{3}{2};Z_3}$
state:
\begin{align*}
&\ \ \ (\frac43,\frac43,2,\frac43,\frac43,2,\frac43,\frac43,0)
\nonumber\\
&=
(2,2,2,2,2,2,2,2,0)- (\frac23,\frac23,0,\frac23,\frac23,0,\frac23,\frac23,0) .
\end{align*}
However, we do not know if a symmetric polynomial described by the
$\v h$ vector $(\frac{12}{9};
\frac43,\frac43,2,\frac43,\frac43,2,\frac43,\frac43,0) $ really
exists or not.

\section{Discussion}

In this paper, we use a local condition -- the pattern of zeros -- to classify
symmetric polynomials of infinity variables.  We find that  symmetric
polynomials of $n$-cluster form (see \eq{PhiFPhiD}, \eq{nClstCond}) can be
labeled by a set in integers $(m;S_2,\cdots,S_n)$.  Those integers must
satisfy the conditions \eq{mSn}, \eq{SanyCnd}, and \eq{PJtot}.

Using the symmetric polynomials labelled by $n$ and
$(m;S_2,\cdots,S_n)$, we have constructed a large class of simple
FQH states.  The constructed FQH states contain both the Laughlin
states and non-Abelian states, such as the Read-Rezayi parafermion
states, the new generalized parafermion states, and some other new
non-Abelian states.  Although, the constructed FQH states are for
bosonic electrons, the bosonic FQH states and the fermionic FQH
states have a simple one-to-one correspondence:
\begin{equation*}
 \Phi_\text{fermion}= \Phi_\text{boson}\prod_{i<j} (z_i-z_j) .
\end{equation*}
We can easily obtain fermionic FQH states from the corresponding
bosonic ones.

We have seen that the ground state wave functions of different Abelian and
non-Abelian fraction quantum Hall states can be characterized by patterns of
zeros $\{S_a\}$.  One may wonder can we use the data $\{S_a\}$ to calculate
various topological properties of the corresponding fraction quantum Hall
state? In \Ref{WWqp}, we will show that many topological properties can indeed
be calculated from $\{S_a\}$, such as the number of possible quasiparticle
types and their quantum numbers.

However, $\{S_a\}$ cannot describe all FQH states.
More complicated ``multi-component'' FQH states, such as
$\nu=2/5$ Abelian FQH state, are not included in our construction.
This suggests that certain non-Abelian state, such as the
parafermion states belong to the same class as the simple
one-component Laughlin states.
Thus, our result can be viewed as a classification of
``one-component'' FQH states, although the precise meaning of
``one-component'' remains to be clarified.  String-net
condensation and the associated tensor category
theory\cite{LWstrnet} provide a fairly complete classification of
non-chiral topological orders in 2 spatial dimensions. We hope the
framework introduced in this paper be a step towards a
classification of chiral topological orders in 2 spatial
dimensions.

\begin{acknowledgments}
We would like to thank M. Freedman and Yue Yu for helpful discussions. 
This research is partially supported by NSF Grant No.  DMR-0706078 (XGW)
and by NSF Grant No. DMS-034772 (ZHW).
\end{acknowledgments}

\appendix

\section{Electrons on a sphere}
\label{EleSph}

To test if the FQH Hamiltonian has an energy gap or not, we need
to put the FQH state on a compact space to avoid the gapless edge
excitations which are always present.\cite{Wedgerev} In this
section, we will discuss how to put a FQH state on a
sphere.\cite{H8305}  We assume that there is a uniform magnetic
field on a sphere with a total $N_\phi$ flux quanta.  The wave
function of one electron in the first Landau level has a
form\cite{H8305}
\begin{equation}
\label{Psiuv}
 \Psi(\th,\vphi)= \sum_{m=0}^{N_\phi} c_m v^{N_\phi-m} u^m
\end{equation}
where $u=\cos(\th/2)e^{i\vphi/2}$ and
$v=\sin(\th/2)e^{-i\vphi/2}$.\footnote{Such a wave function corresponds to a
section of a line bundle on a sphere.  The Chern number of the line bundle is
$N_\phi$.}

We can also use a complex number $z=u/v$ to parameterizes the points
on the sphere.  In term of $z$, the wave function becomes
\begin{align}
\label{Psiz}
\Psi(\th,\vphi) &= e^{-iN_\phi \vphi/2}\sin^{N_\phi}(\th/2)
\sum_{m=0}^{N_\phi} c_m z^m
\nonumber\\
&=\frac{e^{-iN_\phi\vphi/2}}{ (1+|z|^2)^{N_\phi/2}}\sum_{m=0}^{N_\phi} c_m z^m
\end{align}
We see that we can use a polynomial $\Phi(z)$ to describe the wave function of
one electron in the first Landau level:
\begin{equation}
\label{PNphi}
 \Phi(z)=\sum_{m=0}^{N_\phi} c_m z^m
\end{equation}
Here the power of $z$ is equal or less then $N_\phi$.  \eq{Psiuv} and
\eq{Psiz} allow us to go back and forth between the spinor  representation
$\Psi(u,v)$ and the polynomial representation $\Phi(z)$ of the states on
sphere.

Since the polynomials \eq{PNphi}
represent wave functions on a sphere, hence they
form a representation of $SU(2)$ (or $O(3)$)
rotation of the sphere. The dimension
of the representation is $N_\phi+1$. Such a representation
is said to carry an angular momentum
\begin{equation*}
 J=\frac{N_\phi}{2}  .
\end{equation*}
The $SU(2)$ Lie algebra is generated by
\begin{align*}
 L^z=z\prt_z - J,\ \ \ \ \ \ \
 L^-=\prt_z ,\ \ \ \ \ \ \
 L^+=-z^2\prt_z + 2J z ,
\end{align*}
which satisfy
\begin{equation*}
 [L^z, L^\pm]=L^\pm,\ \ \ \ \ \ \ \ [L^+,L^-]=2L^z .
\end{equation*}
Those operators act within the space formed by the
polynomials of the form \eq{PNphi}.
The inner product in the space of the
polynomials is defined through the inner product of the wave functions
$\Psi_1(\th,\vphi)=\frac{e^{-iN_\phi\vphi/2}}{ (1+|z|^2)^{N_\phi/2}}\Phi_1(z)$
and
$\Psi_2(\th,\vphi)=\frac{e^{-iN_\phi\vphi/2}}{ (1+|z|^2)^{N_\phi/2}}\Phi_2(z)$:
\begin{align*}
 \<\Phi_2|\Phi_1\>
&= \int \sin(\th)d\th d\vphi \ \Psi_2^*(\th,\vphi)\Psi_1(\th,\vphi)
\nonumber\\
&= \int 4\cos(\th/2)\sin^3(\th/2)d\frac{\cos(\th/2)}{\sin(\th/2)}
d\vphi \ \Psi_2^*\Psi_1
\nonumber\\
&=\int
\frac{4d^2 z}{(1+|z|^2)^2}
\frac{1}{(1+|z|^2)^{N_\phi}}
\Phi_2^*(z) \Phi_1(z)
\end{align*}

Now let us consider a polynomial of two variables $\Phi(z_1,z_2)$
where the highest power for $z_1$ is $2J_1$ and the highest power
for $z_2$ is $2J_2$ (here $2J_1$ and $2J_2$ are integers).
$\Phi(z_1,z_2)$ can be viewed as a representation of $SU(2)$ too,
where the generators of the $SU(2)$ Lie algebra are given by
\begin{align*}
 L^z&=L^z_1+L^z_2, \ \ \ \ \ \ \ \ L^\pm=L^\pm_1+L^\pm_2,
\nonumber\\
 L^z_1&=z_1\prt_{z_1} - J_1,\ \ \ \
 L^-_1=\prt_{z_1} ,\ \ \ \
 L^+_1=-z_1^2\prt_{z_1} + 2J_1 z_1 ,
\nonumber\\
 L^z_2&=z_2\prt_{z_2} - J_2,\ \ \ \
 L^-_2=\prt_{z_2} ,\ \ \ \
 L^+_2=-z_2^2\prt_{z_2} + 2J_2 z_2 .
\end{align*}
$\Phi(z_1,z_2)$ is not an irreducible representation of $SU(2)$.
It can be decomposed as $\oplus_{J=|J_1-J_2|}^{J_1+J_2} \cH_J$
where $\cH_J$ is an angular-momentum-$J$ representation of
$SU(2)$. We may say that $z_1$ has angular momentum $J_1$ and
$z_2$ has angular momentum $J_2$.  So the angular momenta of
$\Phi(z_1,z_2)$ are those obtained by combining the angular
momentum $J_1$ and the angular momentum $J_2$.

What are the states in the space $\cH_J$? Let $\Phi_{J,m}$,
$m=-J,-J+1,\cdots,J$, be the polynomials in the $\cH_J$ space such that
$\Phi_{J,m}$ is the eigenstate of $L^z$ with eigenvalue $m$.
Let us also introduce $z_\pm = z_1\pm z_2$. We see that
\begin{align*}
L^- &=\prt_{z_1}+\prt_{z_2}= 2\prt_{z_+},
\nonumber\\
L^+ &=
-\frac12 (z_+^2+z_-^2) \prt_{z_+}
-z_+z_- \prt_{z_-} +2J_1 z_1+2J_2z_2 ,
\nonumber\\
L^z&= z_+\prt_{z_+} + z_-\prt_{z_-}-J_1-J_2  .
\end{align*}
From $L^-\Phi_{J,-J}=0$ and $L^z\Phi_{J,-J}=-J\Phi_{J,-J}$, we find that
\begin{equation}
\label{PJmJ}
 \Phi_{J,-J}\propto z_-^{J_1+J_2-J} = (z_1-z_2)^{J_1+J_2-J}
\end{equation}
$\Phi_{J,m}$'s are generated from $\Phi_{J,-J}$ by applying $L^+$'s.  Since
$L^+$ never reduce the power of $z_-$, thus $\Phi_{J,m}$ has a form $
z_-^{J_1+J_2-J}f(z_1,z_2) =(z_1-z_2)^{J_1+J_2-J}f(z_1,z_2) $. This reveals a
close relation between the order of zeros and the angular momentum: \emph{the
polynomial $\Phi(z_1,z_2)$ with angular momentum $J$ must have an order
$J_1+J_2-J$ zero as $z_1\to z_2$}.

On the other hand, let $\Phi_D(z_1,z_2)$ be a polynomial that has
a $D^\text{th}$ order zero as $z_1\to z_2$, \ie $\Phi_D$ has a
form $(z_1-z_2)^D f(z_1,z_2)$. We note that the actions of $L^\pm$
and $L^z$ do not decrease the power of zero as $z_1\to z_2$.  So
the action of $L^\pm$ and $L^z$ can never change $\Phi_D$ to
$\Phi_{J,-J}$ if $J > J_1+J_2-D$. Thus $\Phi_D(z_1,z_2)$ can only
contain angular momenta $J\leq J_1+J_2-D$ which lead to zeros of
order $D$ or more.

We can let $z_1\to z_2$ in $\Phi_D(z_1,z_2)$ and obtain $F(z_2)$ as
\begin{equation*}
 \Phi_D(z_1,z_2)= (z_1-z_2)^D F(z_2) + O[(z_1-z_2)^{D+1}].
\end{equation*}
The minimum $L^z$ eigenvalue for $(z_1-z_2)^D F(z_2)$ is
$-J_1-J_2+D$ which corresponds to $F(z_2)=1$.  After the $SU(2)$
rotations, we generate other polynomials $F(z_2)$ from $F(z_2)=1$.
Those $F(z_2)$ polynomials form an angular momentum $J_1+J_2-D$
representation of $SU(2)$.  Thus \emph{fusing $z_1$ and $z_2$
together produces a variable with an angular momentum $J_1+J_2-D$,
where $J_1$ is the angular momentum of $z_1$, $J_2$ is the angular
momentum of $z_2$, and $D$ is the power of the zeros as $z_1\to
z_2$}.

\section{The relation to conformal field theory}
\label{cft}

It was pointed out that the symmetric polynomial
$\Phi(z_1,\cdots,z_N)$ that describes a FQH state can be written
as an $N$-point correlation function of a certain operator $V_e$
in a conformal field theory (CFT)\cite{MR9162,WWopa,WWHopa}
\begin{equation}
\label{PhiVe}
 \Phi(\{z_i\})=\lim_{z_\infty\to \infty} z_\infty^{2h_N}
\<V(z_\infty)\prod_i V_e(z_i) \>
\end{equation}
To find the corresponding CFT of a symmetric polynomial $\Phi$
described by a pattern of zeros $\{S_a\}$, we will first calculate
the scaling dimension of the operator $V_e$ from $S_a$.

\subsection{The spin of the type-$a$ particles}

Due to the relation between the scaling dimension and the intrinsic spin, we
will first calculate the intrinsic spin of the electron and the $a$-electron
clusters.  From \eq{SaJa}, we see that the angular momentum $J_a$ of the
type-$a$ particle has a form $J_a=aJ-S_a$, where $S_a$ is an intrinsic
property of the type-$a$ particle and is independent of the magnetic flux
$N_\phi=2J$ through the sphere (see \eq{Sbany}).  $S_a$ only depend on the
pattern of the zeros and is called the orbital spin of the type-$a$
particle.\cite{WZspv,WZspvE,Wtoprev}

As discussed in \Ref{WZspvE}, the orbital spin contains two contributions
$S_a=S^\text{sv}_a+h_a$. $S_a^\text{sv}$ comes from the spin vector and $h_a$ is the
intrinsic spin.  The intrinsic spin is related to the statistics of the
particle through the spin-statistics theorem.  The statistics in turn is
related to the scaling dimension (for example, bosons always have integral
scaling dimensions).

To separate the two contributions, we need to identify the contribution from
the spin vector. This can be achieved by noting that the spin vector
contribution is proportional to $a$: $S^\text{sv}_a=C a$. The key is to find the
proportional coefficient $C$.

For this purpose, let us consider $S_{nN_c}$ in \eq{SnNc}.  $S_{nN_c}$ is the
orbital spin for the bound state of $N_c$ type-$n$ particles.  We know that
the type-$n$ particles form a Laughlin state \eq{Pc}.  For a bound state of
$N_c$ type-$n$ particles, its orbital spin $S_{nN_c}$ contains a term linear
in $N_c$ which is the contribution from the spin vector and a term quadratic
in $N_c$ which is the intrinsic spin.  From \eq{SnNc}, we see that the
contribution from the spin vector to $S_{nN_c}$ is
\begin{align*}
S^\text{sv}_{nN_c}&= N_c(S_n-\frac{mn}{2})
\end{align*}
After replacing $N_c$ by $a/n$, we identify the contribution from the spin
vector to $S_{a}$:
\begin{equation}
\label{Ssva}
S^\text{sv}_a=a (\frac{S_n}{n}-\frac{m}{2} )  .
\end{equation}
Thus the intrinsic spin is
\begin{equation}
\label{ispin}
 h_a= S_a-a (\frac{S_n}{n}-\frac{m}{2} ).
\end{equation}
$h_a$ is also the scaling dimension of the operator
$(V_e)^a$.

\subsection{Symmetric polynomial as a correlation in a conformal field theory}

The electron operator $V_e(z)$ in the CFT expression of $\Phi$  \eq{PhiVe}
has a form
\begin{equation*}
 V_e(z) = \psi_1(z) e^{\imth \phi(z)/\sqrt{\nu}}
\end{equation*}
where $e^{\imth \phi/\sqrt{\nu}}$ is the vertex operator in a Gaussian model.
The vertex operator has a scaling dimension $\frac{1}{2\nu}$.  $\psi_1$ is a
simple-current operator,\cite{MR9162,BWnab,WWHopa} \ie
$\psi_1$ satisfy the following fusion relation
\begin{equation*}
 \psi_a\psi_b=\psi_{a+b},\ \ \ \ \ \psi_a\equiv (\psi_1)^a .
\end{equation*}
Such an Abelian fusion rule is closely related to the
unique-fusion condition discussed in section \ref{dPoly}.  If
$\Phi(z_1,\cdots,z_N)$ has an $n$-cluster form, $\psi_1$ satisfies
\begin{equation*}
\psi_n= (\psi_1)^n \sim 1.
\end{equation*}

$\Phi(z_1,\cdots,z_N)$ can be decomposed according to \eq{PhiFPhi}.  The
correlation of the Gausian part $e^{\imth \phi(z)/\sqrt{\nu}}$ produces the
$\Phi_\nu$ part of $\Phi$ and the correlation of the simple-current part
$\psi_1$ produces the $G$ part of $\Phi$.

The intrinsic spin $h_a$ is actually the scaling dimension of the $a^{th}$
power of the electron operator, $V_a\equiv (V_e)^a$.  The scaling dimension of
the Gaussian part $e^{\imth a \phi(z)/\sqrt{\nu}}$ is
$\nu^{-1}\frac{a^2}{2}=\frac{a^2m}{2n}$ and
\begin{equation}
\label{hada}
h^\text{sc}_a=h_a- \frac{a^2m}{2n}
\end{equation}
is the scaling dimension of the simple-current operator
$\psi_a$.

We can obtain the scaling dimension $h_a$ of the operator $V_a$ more directly
without using the concept of spin vector and orbital spin.
First we note that the derived polynomial $P(\{z_i^{(a)}\})$ can be expressed
as a correlation of $V_a(z_i^{(a)})$'s
\begin{equation*}
 P(\{z_i^{(a)}\})=\lim_{z_\infty\to \infty} z_\infty^{2h_N}
\<V(z_\infty)\prod_{i,a} V_a(z_i^{(a)}) \> .
\end{equation*}
The operator product expansion of $V_a$'s determines the powers of zeros in
the correlation function $P(\{z_i^{(a)}\})$.  The $(D_{ab})^{th}$ order
zero as $z_i^{(a)}\to z_i^{(b)}$ implies that
\begin{equation*}
 V_a(z_1) V_b(z_2) \sim (z_1-z_2)^{D_{ab}} V_{a+b}
+ O( (z_1-z_2)^{D_{ab}+1})
\end{equation*}
Thus the scaling dimension of $V_a$ satisfies
\begin{equation}
\label{hhhDel}
h_{a+b} = h_a+h_b +D_{ab}
\end{equation}
From \eq{DabSa}, we see that  $h_a$ and $S_a$ satisfy the same
equation (except the $S_1=0$ condition). Thus
\begin{equation*}
 h_a=S_a - C a
\end{equation*}
for a certain constant $C$.  Since $\psi_{nk}\equiv (\psi_1)^{nk}=1$ and
$V_{kn}=e^{\imth nk\phi/\sqrt{\nu}}$, $h_{nk}$ contains only a term that is
quadratic in $k$.  From \eq{SnNc}, we see that
\begin{align}
\label{Snk}
S_{nk}&=
kS_n+\frac{mnk(k-1)}{2}
\end{align}
Thus we must choose $C$ to cancel the linear $k$ term in  \eq{Snk} to obtain
$h_{nk}$.  We find that $C= \frac{S_n}{n}-\frac{m}{2}$ and
\begin{equation}
\label{ha}
 h_a= S_a- \frac{aS_n}{n}+\frac{am}{2}.
\end{equation}
This allows us to obtain the scaling dimension of the simple current $\psi_a$
which is given by $h^\text{sc}_a$ in \eq{hscaSa}.  We can also express $S_a$ in terms of
$h^\text{sc}_a$ (see \eq{Sahsca}).

We note that CFT requires that $h_a=0$ mod 1, since $V_a$ are bosonic
operators. This requires
\begin{equation}
\label{Snmeven}
 C= \frac{S_n}{n}-\frac{m}{2}=0\text{ mod } 1.
\end{equation}
Such a condition is satisfied if we impose the condition \eq{D3even}.

Let us introduce $V_{-a}\equiv V_a^\dag$ where we have assumed
$\psi_a^\dag=\psi_{n-a}$.  Consider
\begin{align*}
G(z_1,\dots,z_k) &=\<V_{a_1}(z_1)\cdots V_{a_k}(z_k)\>
\nonumber\\ 
&= P(z_1,\cdots,z_k) \prod_{i<j} (z_i-z_j)^{D_{a_i,a_j}}
\end{align*}
where $D_{a,b}=h_{a+b}-h_a-h_b$, $P$ is a polynomial of $z_i$'s,  and $\sum_i
a_i=0$.  The part $\prod_{i<j} (z_i-z_j)^{D_{a,b}}$ reproduces the poles/zeros
of the correlation function as $z_i-z_j\to 0$.  As $z_1\to \infty$,
$\prod_{i<j} (z_i-z_j)^{D_{a,b}}$ behaves as
$ z_1^{\sum_{i=2}^k D_{a_1,a_i}} $.
As a CFT correlation function $G(z_1,\dots,z_k)$ should behave as
$1/z_1^{2h_{a_1}}$ in $z_1\to \infty$ limit.
Thus, the maximal power of $z_1$ in $P(z_1,\cdots,z_k)$ must be
\begin{align*}
\ga_1 &= - \sum_{i=2}^k D_{a_1,a_i} - 2 h_{a_1}  ,
\nonumber\\
 &= (k-3)h_{a_1} + \sum_{i=2}^k h_{a_i} -  \sum_{i=2}^k h_{a_1+a_i},
\end{align*}
where $a_1 =-\sum_{i=2}^k a_i$.
When $k=4$, we have
\begin{equation*}
\ga_1= h_{a_2+a_3+a_4}+h_{a_2}+h_{a_3}+h_{a_4}
-h_{a_2+a_3}
-h_{a_3+a_4}
-h_{a_3+a_2} ,
\end{equation*}
where we have used $h_{-a}=h_a$.  The requirement that $\ga_1\geq 0$ is the
third condition in \eq{SanyCnd}.
%

%

\subsection{Generalized vertex algebra}

To understand the CFT representation of the symmetric polynomial more deeply,
let us consider generalized vertex algebra.\cite{gva}  The CFT formed by the
simple-current operators $\psi_a$ is a special case of a generalized vertex
algebra.

Consider operators $A(z)$, $B(w)$, \etc which form an operator-product-expansion
algebra
\begin{align}
\label{OPE}
& A(z)B(w)=\frac{1}{(z-w)^{\al_{AB}}}
\Big(
[AB]_{\al_{AB}}(w)
\\
&
+(z-w)[AB]_{\al_{AB}-1}(w) 
+(z-w)^2 [AB]_{\al_{AB}-2}(w)  +\cdots
\Big)
\nonumber 
\end{align}
and
\begin{equation}
\label{comAB}
(z-w)^{\al_{AB}} A(z)B(w) =\mu_{AB} (w-z)^{\al_{AB}} B(w) A(z) ,
\end{equation}
where $\mu_{AB}$ is a phase factor.  Here 
$
(z-w)^{\al_{AB}}\equiv |z-w|^{\al_{AB}}\e^{\al_{AB}\imth \th_{zw}}
$
where
$
(z-w)= |z-w|\e^{\imth \th_{zw}}  
$ and $-\pi < \th_{zw} \leq \pi$.
The operator product in \eq{OPE} is assumed to be radial ordered:
$A(z) B(w) \to R[ A(z) B(w)]$ where
\begin{align*}
&\ \ \ \ (z-w)^{\al_{AB}} R[ A(z) B(w)]
\nonumber\\
& =
\begin{cases}
(z-w)^{\al_{AB}} A(z) B(w),  & |z|>|w| , \\
\mu_{AB}(w-z)^{\al_{AB}} B(w) A(z) ,  & |w|>|z| . \\
\end{cases}
\end{align*}
We see that the commutation relation \eq{comAB} ensures that the correlation
functions of $A(z)$ and $B(w)$ are smooth functions.  Let $h_A$, $h_B$ and
$h_{[AB]_{\al_{AB}}}$ be the scaling dimensions of $A$, $B$, and
$[AB]_{\al_{AB}}$, then
\begin{equation*}
 \al_{AB}=h_A+ h_B -h_{[AB]_{\al_{AB}}} .
\end{equation*}

The self consistancy of the vertex algebra
requires $\al_{AB}$'s to statisfy\cite{gva}
\begin{equation}
\label{alABCD}
 \al_{AB}+\al_{AC}-\al_{AD}=0 \text{ mod } 1 ,
\end{equation}
and $\mu_{AB}$'s to statisfy
\begin{equation}
\label{muABCD}
 \mu_{AB}\mu_{AC}=\mu_{AD} (-)^{\al_{AB}+\al_{AC}-\al_{AD}}  ,
\end{equation}
where $D=[BC]_{\al_{BC}}$.

The fusion rule of the simple-current operartors
$\psi_a\psi_b=\psi_{a+b}$ requires that those operators form the following
vertex operator algebra:
\begin{align*}
 \psi_a(z)\psi_b(w) &=\frac{c_{ab}
\Big(\psi_{a+b}(w)+O(z-w) \Big)
}{
(z-w)^{ h^\text{sc}_a +h^\text{sc}_b -h^\text{sc}_{a+b} }
}
\nonumber\\
\psi_a(z)\psi_{-a}(w)&=
\frac{
1+\frac{2h^\text{sc}_a}{c} (z-w)^2 T(w)+O( (z-w)^3)
}
{
(z-w)^{ 2h^\text{sc}_a }
}
\end{align*}
where 
$\psi_{-a}\equiv \psi_{n-a}$ and $\psi_{a+n}\equiv \psi_a$.

We see that the above algebra of simple currents $\psi_a$
is a special case of generalized vertex algebra. We have
\begin{equation*}
\al_{ab}= h^\text{sc}_a +h^\text{sc}_b -h^\text{sc}_{a+b}  .
\end{equation*}
The condition \eq{alABCD} becomes
\begin{equation}
\label{D3mod1}
 \Del_3(a,b,c)=0 \text{ mod } 1  ,
\end{equation}
where we have used \eq{hscaSa1} and \eq{D2D3}.  From \eq{muABCD}, we see that
if $\Del_3(a,b,c)=$ odd for certain choices of $a,b,c$, then $\mu_{ab}$ cannot
be trivial (\ie $\mu_{ab}=1$).  When $\Del_3(a,b,c)=$ even for all $a,b,c$,
then $\mu_{ab}$ can be a trivial solution $\mu_{ab}=1$.  Thus the condition
\eq{D3even} has a special meaning in CFT.

\subsection{The conditions on the $\v h$-vectors}

Due to the one-to-one conrespondence between the $\v S$-vectors and the $\v
h$-vectors [see \eq{hscaSa} and \eq{Sahsca}],
we can translate the conditions \eq{mSn} and \eq{SanyCnd} on the
$\v S$-vectors to some conditions on the $\v h$-vectors.

Note that a $\v h$-vector is specified by $n$, $m$, and
$h^\text{sc}_1,\cdots,h^\text{sc}_{n-1}$.
We extend $h^\text{sc}_1,\cdots,h^\text{sc}_{n-1}$ to $h^\text{sc}_a$
for any integer $a$ by requiring
\begin{equation*}
 h^\text{sc}_0=0,\ \ \ \ \ \ \ h^\text{sc}_a=h^\text{sc}_{a+n} .
\end{equation*}

The conditions \eq{mSn} become
\begin{align}
\label{mhn}
& S_a = h^\text{sc}_a-ah^\text{sc}_1+\frac{a(a-1)m}{2n}=
\text{non-negative integer},
\nonumber\\
& m>0,\ \ \ \ mn=\text{even}, 
\nonumber\\
& 2n h^\text{sc}_1 + m =  0 \text{ mod } n.
\end{align}

From $2n h^\text{sc}_1 + m =  0 \text{ mod } n$, we see  that $2n h^\text{sc}_1$ is an
integer.  From $2n h^\text{sc}_a-a (2n h^\text{sc}_1)+a(a-1)m=$ even integer, we see that
$2nh^\text{sc}_a$ are always integers.  Also $2nh^\text{sc}_{2a}$ are always even
integers and $2nh^\text{sc}_{2a+1}$ are either all even or all odd.  Since
$h^\text{sc}_n=0$, thus when $n=$ odd, $2nh^\text{sc}_a$ are all even.  Only when
$n=$ even, can $2nh^\text{sc}_{2a+1}$ either be all even or all odd.

To generate sets of $h^\text{sc}_a$ that satisfy the above conditions,
we will use \eq{hscaSa}.
Setting $a=1$ in \eq{hscaSa}, we get
$2n h^\text{sc}_1=m(n-1)-2 S_n$.
We see that when $n=$ odd, $ 2n h^\text{sc}_1=$ even.  When $n=$  even, $2n
h^\text{sc}_1=$ even when $m=$ even and $2n h^\text{sc}_1=$ odd when $m=$ odd.
We also see that $S_n\leq m(n-1)/2$ which implies that.
$S_a\leq m(n-1)/2$ for $a=2,3,\cdots,n$.

The conditions \eq{SanyCnd} become
\begin{align}
\label{hanyCnd}
& n h^\text{sc}_{2a} -2 a n h^\text{sc}_1+ ma(2a-1)=0 \text{ mod } 2n,
\nonumber\\
& h^\text{sc}_{a+b}-h^\text{sc}_a-h^\text{sc}_b \geq  - \frac{abm}{n}, 
\\
& h^\text{sc}_{a+b+c} -h^\text{sc}_{a+b} -h^\text{sc}_{b+c} -h^\text{sc}_{a+c}
+h^\text{sc}_a +h^\text{sc}_b +h^\text{sc}_c \geq 0.  
\nonumber 
\end{align}
The condition \eq{D3even} becomes
\begin{align}
\label{D3evenh}
& h^\text{sc}_{a+b+c} -h^\text{sc}_{a+b} -h^\text{sc}_{b+c} -h^\text{sc}_{a+c}
+h^\text{sc}_a +h^\text{sc}_b +h^\text{sc}_c =\text{even}.  
\end{align}

The first condition in
\eq{mhn} and the second condition in \eq{hanyCnd} imply that
\begin{align}
& h^\text{sc}_{a+b}-h^\text{sc}_a-h^\text{sc}_b + \frac{abm}{n} \geq 0, 
\nonumber\\
& h^\text{sc}_{a+b}-h^\text{sc}_a-h^\text{sc}_b + \frac{abm}{n} = 0 \text{ mod
} 1, 
\end{align}
which are part of defining conditions of parafermion CFT if $m=2$.
So the pattern of zeros of symmetric polynomial may have a natural relation to
parafermion CFT.

%



\begin{thebibliography}{46}
\expandafter\ifx\csname natexlab\endcsname\relax\def\natexlab#1{#1}\fi
\expandafter\ifx\csname bibnamefont\endcsname\relax
  \def\bibnamefont#1{#1}\fi
\expandafter\ifx\csname bibfnamefont\endcsname\relax
  \def\bibfnamefont#1{#1}\fi
\expandafter\ifx\csname citenamefont\endcsname\relax
  \def\citenamefont#1{#1}\fi
\expandafter\ifx\csname url\endcsname\relax
  \def\url#1{\texttt{#1}}\fi
\expandafter\ifx\csname urlprefix\endcsname\relax\def\urlprefix{URL }\fi
\providecommand{\bibinfo}[2]{#2}
\providecommand{\eprint}[2][]{\url{#2}}

\bibitem[{\citenamefont{Landau}(1937)}]{L3726}
\bibinfo{author}{\bibfnamefont{L.~D.} \bibnamefont{Landau}},
  \bibinfo{journal}{Phys. Z. Sowjetunion} \textbf{\bibinfo{volume}{11}},
  \bibinfo{pages}{26} (\bibinfo{year}{1937}).

\bibitem[{\citenamefont{Tsui et~al.}(1982)\citenamefont{Tsui, Stormer, and
  Gossard}}]{TSG8259}
\bibinfo{author}{\bibfnamefont{D.~C.} \bibnamefont{Tsui}},
  \bibinfo{author}{\bibfnamefont{H.~L.} \bibnamefont{Stormer}},
  \bibnamefont{and} \bibinfo{author}{\bibfnamefont{A.~C.}
  \bibnamefont{Gossard}}, \bibinfo{journal}{Phys. Rev. Lett.}
  \textbf{\bibinfo{volume}{48}}, \bibinfo{pages}{1559} (\bibinfo{year}{1982}).

\bibitem[{\citenamefont{Laughlin}(1983)}]{L8395}
\bibinfo{author}{\bibfnamefont{R.~B.} \bibnamefont{Laughlin}},
  \bibinfo{journal}{Phys. Rev. Lett.} \textbf{\bibinfo{volume}{50}},
  \bibinfo{pages}{1395} (\bibinfo{year}{1983}).

\bibitem[{\citenamefont{Wen}(1990)}]{Wrig}
\bibinfo{author}{\bibfnamefont{X.-G.} \bibnamefont{Wen}},
  \bibinfo{journal}{Int. J. Mod. Phys. B} \textbf{\bibinfo{volume}{4}},
  \bibinfo{pages}{239} (\bibinfo{year}{1990}).

\bibitem[{\citenamefont{Wen and Niu}(1990)}]{WNtop}
\bibinfo{author}{\bibfnamefont{X.-G.} \bibnamefont{Wen}} \bibnamefont{and}
  \bibinfo{author}{\bibfnamefont{Q.}~\bibnamefont{Niu}},
  \bibinfo{journal}{Phys. Rev. B} \textbf{\bibinfo{volume}{41}},
  \bibinfo{pages}{9377} (\bibinfo{year}{1990}).

\bibitem[{\citenamefont{Kitaev and Preskill}(2006)}]{KP0604}
\bibinfo{author}{\bibfnamefont{A.}~\bibnamefont{Kitaev}} \bibnamefont{and}
  \bibinfo{author}{\bibfnamefont{J.}~\bibnamefont{Preskill}},
  \bibinfo{journal}{Phys. Rev. Lett.} \textbf{\bibinfo{volume}{96}},
  \bibinfo{pages}{110404} (\bibinfo{year}{2006}).

\bibitem[{\citenamefont{Levin and Wen}(2006)}]{LWtopent}
\bibinfo{author}{\bibfnamefont{M.}~\bibnamefont{Levin}} \bibnamefont{and}
  \bibinfo{author}{\bibfnamefont{X.-G.} \bibnamefont{Wen}},
  \bibinfo{journal}{Phys. Rev. Lett.} \textbf{\bibinfo{volume}{96}},
  \bibinfo{pages}{110405} (\bibinfo{year}{2006}).

\bibitem[{\citenamefont{Kalmeyer and Laughlin}(1987)}]{KL8795}
\bibinfo{author}{\bibfnamefont{V.}~\bibnamefont{Kalmeyer}} \bibnamefont{and}
  \bibinfo{author}{\bibfnamefont{R.~B.} \bibnamefont{Laughlin}},
  \bibinfo{journal}{Phys. Rev. Lett.} \textbf{\bibinfo{volume}{59}},
  \bibinfo{pages}{2095} (\bibinfo{year}{1987}).

\bibitem[{\citenamefont{Rokhsar and Kivelson}(1988)}]{RK8876}
\bibinfo{author}{\bibfnamefont{D.~S.} \bibnamefont{Rokhsar}} \bibnamefont{and}
  \bibinfo{author}{\bibfnamefont{S.~A.} \bibnamefont{Kivelson}},
  \bibinfo{journal}{Phys. Rev. Lett.} \textbf{\bibinfo{volume}{61}},
  \bibinfo{pages}{2376} (\bibinfo{year}{1988}).

\bibitem[{\citenamefont{Affleck and Marston}(1988)}]{AM8874}
\bibinfo{author}{\bibfnamefont{I.}~\bibnamefont{Affleck}} \bibnamefont{and}
  \bibinfo{author}{\bibfnamefont{J.~B.} \bibnamefont{Marston}},
  \bibinfo{journal}{Phys. Rev. B} \textbf{\bibinfo{volume}{37}},
  \bibinfo{pages}{3774} (\bibinfo{year}{1988}).

\bibitem[{\citenamefont{Wen et~al.}(1989)\citenamefont{Wen, Wilczek, and
  Zee}}]{WWZcsp}
\bibinfo{author}{\bibfnamefont{X.-G.} \bibnamefont{Wen}},
  \bibinfo{author}{\bibfnamefont{F.}~\bibnamefont{Wilczek}}, \bibnamefont{and}
  \bibinfo{author}{\bibfnamefont{A.}~\bibnamefont{Zee}},
  \bibinfo{journal}{Phys. Rev. B} \textbf{\bibinfo{volume}{39}},
  \bibinfo{pages}{11413} (\bibinfo{year}{1989}).

\bibitem[{\citenamefont{Read and Sachdev}(1991)}]{RS9173}
\bibinfo{author}{\bibfnamefont{N.}~\bibnamefont{Read}} \bibnamefont{and}
  \bibinfo{author}{\bibfnamefont{S.}~\bibnamefont{Sachdev}},
  \bibinfo{journal}{Phys. Rev. Lett.} \textbf{\bibinfo{volume}{66}},
  \bibinfo{pages}{1773} (\bibinfo{year}{1991}).

\bibitem[{\citenamefont{Wen}(1991{\natexlab{a}})}]{Wsrvb}
\bibinfo{author}{\bibfnamefont{X.-G.} \bibnamefont{Wen}},
  \bibinfo{journal}{Phys. Rev. B} \textbf{\bibinfo{volume}{44}},
  \bibinfo{pages}{2664} (\bibinfo{year}{1991}{\natexlab{a}}).

\bibitem[{\citenamefont{Moore and Read}(1991)}]{MR9162}
\bibinfo{author}{\bibfnamefont{G.}~\bibnamefont{Moore}} \bibnamefont{and}
  \bibinfo{author}{\bibfnamefont{N.}~\bibnamefont{Read}},
  \bibinfo{journal}{Nucl. Phys. B} \textbf{\bibinfo{volume}{360}},
  \bibinfo{pages}{362} (\bibinfo{year}{1991}).

\bibitem[{\citenamefont{Wen}(1991{\natexlab{b}})}]{Wnab}
\bibinfo{author}{\bibfnamefont{X.-G.} \bibnamefont{Wen}},
  \bibinfo{journal}{Phys. Rev. Lett.} \textbf{\bibinfo{volume}{66}},
  \bibinfo{pages}{802} (\bibinfo{year}{1991}{\natexlab{b}}).

\bibitem[{\citenamefont{Blok and Wen}(1990)}]{BWkmat1}
\bibinfo{author}{\bibfnamefont{B.}~\bibnamefont{Blok}} \bibnamefont{and}
  \bibinfo{author}{\bibfnamefont{X.-G.} \bibnamefont{Wen}},
  \bibinfo{journal}{Phys. Rev. B} \textbf{\bibinfo{volume}{42}},
  \bibinfo{pages}{8133} (\bibinfo{year}{1990}).

\bibitem[{\citenamefont{Read}(1990)}]{R9002}
\bibinfo{author}{\bibfnamefont{N.}~\bibnamefont{Read}}, \bibinfo{journal}{Phys.
  Rev. Lett.} \textbf{\bibinfo{volume}{65}}, \bibinfo{pages}{1502}
  (\bibinfo{year}{1990}).

\bibitem[{\citenamefont{Fr{\"o}hlich and Kerler}(1991)}]{FK9169}
\bibinfo{author}{\bibfnamefont{J.}~\bibnamefont{Fr{\"o}hlich}}
  \bibnamefont{and} \bibinfo{author}{\bibfnamefont{T.}~\bibnamefont{Kerler}},
  \bibinfo{journal}{Nucl. Phys. B} \textbf{\bibinfo{volume}{354}},
  \bibinfo{pages}{369} (\bibinfo{year}{1991}).

\bibitem[{\citenamefont{Read and Rezayi}(1999)}]{RR9984}
\bibinfo{author}{\bibfnamefont{N.}~\bibnamefont{Read}} \bibnamefont{and}
  \bibinfo{author}{\bibfnamefont{E.}~\bibnamefont{Rezayi}},
  \bibinfo{journal}{Phys.Rev. B} \textbf{\bibinfo{volume}{59}},
  \bibinfo{pages}{8084} (\bibinfo{year}{1999}).

\bibitem[{\citenamefont{Misguich et~al.}(1999)\citenamefont{Misguich,
  Lhuillier, Bernu, and Waldtmann}}]{MLB9964}
\bibinfo{author}{\bibfnamefont{G.}~\bibnamefont{Misguich}},
  \bibinfo{author}{\bibfnamefont{C.}~\bibnamefont{Lhuillier}},
  \bibinfo{author}{\bibfnamefont{B.}~\bibnamefont{Bernu}}, \bibnamefont{and}
  \bibinfo{author}{\bibfnamefont{C.}~\bibnamefont{Waldtmann}},
  \bibinfo{journal}{Phys. Rev. B} \textbf{\bibinfo{volume}{60}},
  \bibinfo{pages}{1064} (\bibinfo{year}{1999}).

\bibitem[{\citenamefont{Moessner and Sondhi}(2001)}]{MS0181}
\bibinfo{author}{\bibfnamefont{R.}~\bibnamefont{Moessner}} \bibnamefont{and}
  \bibinfo{author}{\bibfnamefont{S.~L.} \bibnamefont{Sondhi}},
  \bibinfo{journal}{Phys. Rev. Lett.} \textbf{\bibinfo{volume}{86}},
  \bibinfo{pages}{1881} (\bibinfo{year}{2001}).

\bibitem[{\citenamefont{Wen}(2002)}]{Wqoslpub}
\bibinfo{author}{\bibfnamefont{X.-G.} \bibnamefont{Wen}},
  \bibinfo{journal}{Phys. Rev. B} \textbf{\bibinfo{volume}{65}},
  \bibinfo{pages}{165113} (\bibinfo{year}{2002}).

\bibitem[{\citenamefont{Kitaev}(2003)}]{K032}
\bibinfo{author}{\bibfnamefont{A.~Y.} \bibnamefont{Kitaev}},
  \bibinfo{journal}{Ann. Phys. (N.Y.)} \textbf{\bibinfo{volume}{303}},
  \bibinfo{pages}{2} (\bibinfo{year}{2003}).

\bibitem[{\citenamefont{Freedman et~al.}(2003)\citenamefont{Freedman, Kitaev,
  Larsen, and Wang}}]{FKL0331}
\bibinfo{author}{\bibfnamefont{M.~H.} \bibnamefont{Freedman}},
  \bibinfo{author}{\bibfnamefont{A.}~\bibnamefont{Kitaev}},
  \bibinfo{author}{\bibfnamefont{M.~J.} \bibnamefont{Larsen}},
  \bibnamefont{and} \bibinfo{author}{\bibfnamefont{Z.}~\bibnamefont{Wang}},
  \bibinfo{journal}{Bull. Amer. Math. Soc.} \textbf{\bibinfo{volume}{40}},
  \bibinfo{pages}{31} (\bibinfo{year}{2003}).

\bibitem[{\citenamefont{Freedman et~al.}(2004)\citenamefont{Freedman, Nayak,
  Shtengel, Walker, and Wang}}]{FNS0428}
\bibinfo{author}{\bibfnamefont{M.}~\bibnamefont{Freedman}},
  \bibinfo{author}{\bibfnamefont{C.}~\bibnamefont{Nayak}},
  \bibinfo{author}{\bibfnamefont{K.}~\bibnamefont{Shtengel}},
  \bibinfo{author}{\bibfnamefont{K.}~\bibnamefont{Walker}}, \bibnamefont{and}
  \bibinfo{author}{\bibfnamefont{Z.}~\bibnamefont{Wang}},
  \bibinfo{journal}{Ann. Phys. (NY)} \textbf{\bibinfo{volume}{310}},
  \bibinfo{pages}{428} (\bibinfo{year}{2004}).

\bibitem[{\citenamefont{Levin and Wen}(2005)}]{LWstrnet}
\bibinfo{author}{\bibfnamefont{M.}~\bibnamefont{Levin}} \bibnamefont{and}
  \bibinfo{author}{\bibfnamefont{X.-G.} \bibnamefont{Wen}},
  \bibinfo{journal}{Phys. Rev. B} \textbf{\bibinfo{volume}{71}},
  \bibinfo{pages}{045110} (\bibinfo{year}{2005}).

\bibitem[{\citenamefont{Kitaev}(2005)}]{K0538}
\bibinfo{author}{\bibfnamefont{A.}~\bibnamefont{Kitaev}},
  \bibinfo{journal}{cond-mat/0506438}
  (\bibinfo{year}{2005}).

\bibitem[{\citenamefont{Read}(2006)}]{R0678}
\bibinfo{author}{\bibfnamefont{N.}~\bibnamefont{Read}},
  \bibinfo{journal}{cond-mat/0601678}
  (\bibinfo{year}{2006}).

\bibitem[{\citenamefont{Wen}(1995)}]{Wtoprev}
\bibinfo{author}{\bibfnamefont{X.-G.} \bibnamefont{Wen}},
  \bibinfo{journal}{Advances in Physics} \textbf{\bibinfo{volume}{44}},
  \bibinfo{pages}{405} (\bibinfo{year}{1995}).

\bibitem[{\citenamefont{Wen}(2004)}]{Wen04}
\bibinfo{author}{\bibfnamefont{X.-G.} \bibnamefont{Wen}},
  \emph{\bibinfo{title}{Quantum Field Theory of Many-Body Systems -- From the
  Origin of Sound to an Origin of Light and Electrons}}
  (\bibinfo{publisher}{Oxford Univ. Press}, \bibinfo{address}{Oxford},
  \bibinfo{year}{2004}).

\bibitem[{\citenamefont{Seidel and Lee}(2006)}]{SL0604}
\bibinfo{author}{\bibfnamefont{A.}~\bibnamefont{Seidel}} \bibnamefont{and}
  \bibinfo{author}{\bibfnamefont{D.-H.} \bibnamefont{Lee}},
  \bibinfo{journal}{Phys. Rev. Lett.} \textbf{\bibinfo{volume}{97}},
  \bibinfo{pages}{056804} (\bibinfo{year}{2006}).

\bibitem[{\citenamefont{Bergholtz et~al.}(2006)\citenamefont{Bergholtz,
  Kailasvuori, Wikberg, Hansson, and Karlhede}}]{BKW0608}
\bibinfo{author}{\bibfnamefont{E.}~\bibnamefont{Bergholtz}},
  \bibinfo{author}{\bibfnamefont{J.}~\bibnamefont{Kailasvuori}},
  \bibinfo{author}{\bibfnamefont{E.}~\bibnamefont{Wikberg}},
  \bibinfo{author}{\bibfnamefont{T.}~\bibnamefont{Hansson}}, \bibnamefont{and}
  \bibinfo{author}{\bibfnamefont{A.}~\bibnamefont{Karlhede}},
  \bibinfo{journal}{Phys. Rev. B} \textbf{\bibinfo{volume}{74}},
  \bibinfo{pages}{081308} (\bibinfo{year}{2006}).

\bibitem[{\citenamefont{Seidel and Yang}(2008)}]{SY0802}
\bibinfo{author}{\bibfnamefont{A.}~\bibnamefont{Seidel}} \bibnamefont{and}
  \bibinfo{author}{\bibfnamefont{K.}~\bibnamefont{Yang}},
  \bibinfo{journal}{arXiv:0801.2402} (\bibinfo{year}{2008}).

\bibitem[{\citenamefont{Bernevig and Haldane}(2007{\natexlab{a}})}]{BH0737}
\bibinfo{author}{\bibfnamefont{B.~A.} \bibnamefont{Bernevig}} \bibnamefont{and}
  \bibinfo{author}{\bibfnamefont{F.~D.~M.} \bibnamefont{Haldane}},
  \bibinfo{journal}{arXiv:0707.3637} (\bibinfo{year}{2007}{\natexlab{a}}).

\bibitem[{\citenamefont{Bernevig and Haldane}(2007{\natexlab{b}})}]{BH0762}
\bibinfo{author}{\bibfnamefont{B.~A.} \bibnamefont{Bernevig}} \bibnamefont{and}
  \bibinfo{author}{\bibfnamefont{F.}~\bibnamefont{Haldane}},
  \bibinfo{journal}{arXiv:0711.3062} (\bibinfo{year}{2007}{\natexlab{b}}).

\bibitem[{\citenamefont{Haldane}(1983)}]{H8305}
\bibinfo{author}{\bibfnamefont{F.~D.~M.} \bibnamefont{Haldane}},
  \bibinfo{journal}{Phys. Rev. Lett.} \textbf{\bibinfo{volume}{51}},
  \bibinfo{pages}{605} (\bibinfo{year}{1983}).

\bibitem[{\citenamefont{Greiter et~al.}(1991)\citenamefont{Greiter, Wen, and
  Wilczek}}]{GWW9105}
\bibinfo{author}{\bibfnamefont{M.}~\bibnamefont{Greiter}},
  \bibinfo{author}{\bibfnamefont{X.-G.} \bibnamefont{Wen}}, \bibnamefont{and}
  \bibinfo{author}{\bibfnamefont{F.}~\bibnamefont{Wilczek}},
  \bibinfo{journal}{Phys. Rev. Lett.} \textbf{\bibinfo{volume}{66}},
  \bibinfo{pages}{3205} (\bibinfo{year}{1991}).

\bibitem[{\citenamefont{Wen}(1993)}]{Wnabhalf}
\bibinfo{author}{\bibfnamefont{X.-G.} \bibnamefont{Wen}},
  \bibinfo{journal}{Phys. Rev. Lett.} \textbf{\bibinfo{volume}{70}},
  \bibinfo{pages}{355} (\bibinfo{year}{1993}).

\bibitem[{\citenamefont{Haldane}(2006)}]{H0633}
\bibinfo{author}{\bibfnamefont{F.}~\bibnamefont{Haldane}},
  \bibinfo{journal}{Bull. Am. Phys. Soc.} \textbf{\bibinfo{volume}{51}},
  \bibinfo{pages}{633} (\bibinfo{year}{2006}).

\bibitem[{\citenamefont{Blok and Wen}(1992)}]{BWnab}
\bibinfo{author}{\bibfnamefont{B.}~\bibnamefont{Blok}} \bibnamefont{and}
  \bibinfo{author}{\bibfnamefont{X.-G.} \bibnamefont{Wen}},
  \bibinfo{journal}{Nucl. Phys. B} \textbf{\bibinfo{volume}{374}},
  \bibinfo{pages}{615} (\bibinfo{year}{1992}).

\bibitem[{\citenamefont{Wen and Wu}(1994)}]{WWopa}
\bibinfo{author}{\bibfnamefont{X.-G.} \bibnamefont{Wen}} \bibnamefont{and}
  \bibinfo{author}{\bibfnamefont{Y.-S.} \bibnamefont{Wu}},
  \bibinfo{journal}{Nucl. Phys. B} \textbf{\bibinfo{volume}{419}},
  \bibinfo{pages}{455} (\bibinfo{year}{1994}).

\bibitem[{\citenamefont{Wen et~al.}(1994)\citenamefont{Wen, Wu, and
  Hatsugai}}]{WWHopa}
\bibinfo{author}{\bibfnamefont{X.-G.} \bibnamefont{Wen}},
  \bibinfo{author}{\bibfnamefont{Y.-S.} \bibnamefont{Wu}}, \bibnamefont{and}
  \bibinfo{author}{\bibfnamefont{Y.}~\bibnamefont{Hatsugai}},
  \bibinfo{journal}{Nucl. Phys. B} \textbf{\bibinfo{volume}{422}},
  \bibinfo{pages}{476} (\bibinfo{year}{1994}).

\bibitem[{\citenamefont{Read}(2006)}]{R0634}
\bibinfo{author}{\bibfnamefont{N.}~\bibnamefont{Read}}, \bibinfo{journal}{Phys.
  Rev. B} \textbf{\bibinfo{volume}{73}}, \bibinfo{pages}{245334}
  (\bibinfo{year}{2006}).

\bibitem[{\citenamefont{Simon et~al.}(2007)\citenamefont{Simon, Rezayi, and
  Cooper}}]{SRC0760}
\bibinfo{author}{\bibfnamefont{S.~H.} \bibnamefont{Simon}},
  \bibinfo{author}{\bibfnamefont{E.~H.} \bibnamefont{Rezayi}},
  \bibnamefont{and} \bibinfo{author}{\bibfnamefont{N.~R.}
  \bibnamefont{Cooper}}, \bibinfo{journal}{cond-mat/0701260} (\bibinfo{year}{2007}).

\bibitem[{\citenamefont{Wen}(1992)}]{Wedgerev}
\bibinfo{author}{\bibfnamefont{X.-G.} \bibnamefont{Wen}},
  \bibinfo{journal}{Int. J. Mod. Phys. B} \textbf{\bibinfo{volume}{6}},
  \bibinfo{pages}{1711} (\bibinfo{year}{1992}).

\bibitem[{\citenamefont{Wen and Zee}(1992{\natexlab{a}})}]{WZspv}
\bibinfo{author}{\bibfnamefont{X.-G.} \bibnamefont{Wen}} \bibnamefont{and}
  \bibinfo{author}{\bibfnamefont{A.}~\bibnamefont{Zee}},
  \bibinfo{journal}{Phys. Rev. Lett.} \textbf{\bibinfo{volume}{69}},
  \bibinfo{pages}{953} (\bibinfo{year}{1992}{\natexlab{a}}).

\bibitem[{\citenamefont{Wen and Zee}(1992{\natexlab{b}})}]{WZspvE}
\bibinfo{author}{\bibfnamefont{X.-G.} \bibnamefont{Wen}} \bibnamefont{and}
  \bibinfo{author}{\bibfnamefont{A.}~\bibnamefont{Zee}},
  \bibinfo{journal}{Phys. Rev. Lett.} \textbf{\bibinfo{volume}{69}},
  \bibinfo{pages}{3000} (\bibinfo{year}{1992}{\natexlab{b}}).

\bibitem{YueYu} Yue Yu, unpublished.

\bibitem{WWqp} 
X.-G. Wen and Z. Wang, arXiv:0803.1016 (2008)

\bibitem{gva} Boris Noyvert,
Journal of High Energy Physics,
\textbf{2}, 74 (2007); hep-th/0602273.


\end{thebibliography}

\end{document}